# The Sun's Fast Dynamo Action

## D. V. Sarafopoulos

*Department of Electrical and Computer Engineering,*

*Democritus University of Thrace, Xanthi, Greece*

*sarafo@ee.duth.gr*

# Abstract

We provide a synthesis model demonstrating the "fast dynamo" action of the Sun. The latter is essentially accomplished via two toroidal structures presumably formed in the tachocline and placed symmetrically with respect to the equatorial plane. The two tori are characterized by several prominent key-properties as follows: First, ***in each "Torus" a surplus charge is entrapped for approximately the time period of an 11-year sunspot cycle; for the next cycle the charge changes sign***. Second, the net charge of Torus, moving with the solar rotational speed, generates a huge toroidal current which, in turn, builds up an intense poloidal magnetic field. Third, each Torus is placed at a specific distance from the Radiative Zone, so that the rotational speed ($u_\varphi$) of an entrapped charge carrier equals the local propagation velocity for an electromagnetic disturbance ($c'$). Thus, ***<u>two Torus electrons satisfy the condition that the repulsive electrostatic force equals the attractive magnetic force</u>*** caused from the two elementary currents. Fourth, the surplus charge can steadily and unanticipatedly increase; electrons are systematically attracted inwards or repealed outwards via the poloidal field. The charges remain without any "Debye shielding" action in the Torus-core region, while they demonstrate a ***"Debye anti-shielding effect"*** closer to the Radiative Zone (where $u_\varphi > c'$). ***Thus, the Torus charge "moves" with zero resistivity and velocity $\sim$1500 m/s***, i.e. the Sun's rotational speed. ***<u>The Torus core region behaves like a gigantic "superconductor" at the extremely high temperatures of tachocline.</u>*** Fifth, the tori move equatorward drifting on a surface being an ellipsoid by revolution on which the condition $u_\varphi = c'$ is satisfied. At the end of an 11–year cycle, the tori finally disintegrate close to the equatorial plane, given that their toroidal magnetic field lines are oppositely directed. Moreover, we present a preliminary 3D solar circuit, for the overall 22-year cycle, with the ability to reverse the magnetic field. If the suggested model is accepted as a workable solution, then many longstanding unresolved questions concerning the powerful CMEs, the flares, the electron acceleration mechanism and the stellar dynamo could be readily addressed.

**Key words:** fast dynamo, slow dynamo, solar dynamo, stellar dynamo, CME, solar flare, solar 3D electric circuit, Debye anti-shielding effect, electron acceleration, solar cycle, superconductivity, superconductivity at extremely high temperatures, Geodynamo.

# *1. Introduction*

In general, the "dynamo" and "reconnection" effects, in space plasma physics, are two fundamental and mutually related topics with crucial importance. This author, in the past, is mainly engaged with the magnetic field reconnection process associated with in situ satellite measurements (e.g., [1, 4]). In this work, we suggest a "fast dynamo action" being potentially at work in the Sun's case and producing the main features related to successive 11-year sunspot cycles (e.g., [5]). The model is completely distanced from any MHD approach; the needed "zero resistivity" does not result, as it is commonly anticipated, through a convection "velocity field". Instead, an unusual physical mechanism can produce a surplus charge within the "two toroidal structures" formed in the Sun's tachocline. And the net charges, rotating with the Sun's rotational speed, produce very intense azimuthal currents. That is, we essentially suggest that ***two superconductive toroidal structures are formed and finally collapsed in each 11-year cycle***, inside the Sun. The resulting 3D solar circuit composed of azimuthal and meridional currents (generated in the space between the Radiative Zone and photosphere) are responsible for launching the activity cycle in the Sun. We shall thoroughly exhibit all these aspects; however, at present a theory prevails regarding the Sun's dynamo and the generation and variation of the magnetic field of the Sun. For a better understanding of the model which is proposed in this paper, the evolution of this theory will be described in the following.

In solar or planetary physics, the **dynamo theory** proposes a mechanism by which a celestial body such as Earth, the Sun or a star generates a magnetic field. It is typically repeated that the "dynamo theory describes the process through which a rotating, convecting, and electrically conducting fluid acts to maintain a magnetic field. This theory is used to explain the presence of anomalously long-lived magnetic fields in astrophysical bodies" (e.g., [6] and all the other articles included in the same book "Lectures on solar and planetary dynamos"). However, it is commonly stressed that the detailed mechanism of the solar dynamo is at present not known in detail and is therefore the subject of current research (e.g., [7]).

Traditionally, dynamos are divided into "kinematic dynamos" and "nonlinear dynamos" or "hydromagnetic dynamos" which emphasize the importance of hydromagnetic interactions ([8, 14]). In kinematic dynamo theory the velocity field is prescribed, instead of being a dynamic variable. The most functional feature of kinematic dynamo theory is that it can be used to test whether a velocity field is or is not capable of dynamo action. By applying a certain (steady-state) velocity field to a small magnetic field, it can be determined through observation whether the magnetic field tends to grow or not in reaction to the applied flow. If the magnetic field does grow, then the system is either capable to perform dynamo action or constitutes a dynamo itself.

Using Maxwell's equations simultaneously with the curl of Ohm's Law, one can derive what is basically the linear MHD kinematic dynamo (or induction) equation for magnetic fields (**B**), which can be done when assuming that the magnetic field is independent from the velocity field. The **B** and **u** are directly related

$$\frac{\partial \mathbf{B}}{\partial t} = \nabla \times (\mathbf{u} \times \mathbf{B}) + \frac{\eta}{\mu} \nabla^2 \mathbf{B}$$

where **u, B,** t, η and μ are the fluid velocity, magnetic field, time, electrical resistivity and permeability, respectively.

The kinematic approximation becomes invalid when the magnetic field becomes strong enough to affect the fluid motions. In that case (of "backreaction") the velocity field becomes affected by the Lorentz force, and so the induction equation is no longer linear in the magnetic field. In most cases this leads to a quenching of the amplitude of the dynamo. Such dynamos are sometimes also referred to as hydromagnetic (self-consistent) dynamos. Virtually all dynamos in astrophysics and geophysics are hydromagnetic dynamos.

The equations for a solar dynamo are enormously difficult to solve, and the realism of the solutions is mainly restricted by computer power. For decades, theorists were confined to kinematic dynamo models, in which the fluid motion parameters are chosen in advance and the effect on the magnetic field is calculated. Kinematic dynamo theory was mainly a matter of trying different flow geometries and seeing whether they could fulfil the dynamo conditions.

The kinematic dynamo models are subdivided in laminar and turbulent types. With respect to the growth rate of the magnetic field, the laminar dynamos are named slow or fast. A homopolar disk generator (which comprises an electrically conductive disc rotating in a plane perpendicular to a uniform static magnetic field) is an example of a "slow dynamo" since its growth-rate depends on the resistivity. A potential difference is created between the center of the disc and the rim; thus, slow dynamos can be manufactured and undoubtedly exist. Furthermore, any dynamo, which depends on the motion of a rigid conductor for its operation, fulfils the conditions of a slow dynamo. In the perfectly conducting limit, the magnetic flux linking the conductor could never change, so there would be no increase of the magnetic field. Consequently, in this perfectly conducting limit, the magnetic field-lines are always considered frozen into the MHD fluid. However, it is generally believed that "fast dynamo action" is a possibility for an MHD fluid. And this belief is based on the fact that an MHD fluid is a non-rigid body, and thus, its motion possesses degrees of freedom not accessible to rigid conductors ([15]).

If the fluid is incompressible, then the stretching of field-lines implies a proportionate intensification of the field-strength. The simplest heuristic fast dynamo, first described by Vainshtein and Zel'dovich [16], is based on this effect. A magnetic flux-tube can be doubled in intensity by placing it around a "stretch-twist-fold cycle". The doubling time for this process clearly does not depend on the resistivity: in this sense, the ohmic losses do not damp the fluid motion and the dynamo turns out to be a fast dynamo. However, under repeated application of this cycle the magnetic field develops an increasingly fine-scale structure and both the **u** and **B** fields eventually become chaotic.

It should be emphasized that at present, the physical existence of fast dynamos has not been conclusively established, since most of the literature on this subject is based on mathematical paradigms rather than actual solutions of the dynamo equation. It should be noted, however, that ***the need for fast dynamo solutions is fairly acute, especially in stellar dynamo theory***. For instance, for the Sun the ohmic decay time is about $10^{12}$ years, whereas the reversal time for the solar magnetic field is only 11 years. Thus, it can hardly be assumed that resistivity plays any significant role in the solar dynamo.

The solar dynamo involves an intricate interplay of complex processes occurring over a wide range of spatial and temporal scales. Consequently, global convection simulations are a long way from making detailed comparisons with photosphere and coronal observations of magnetic activity. ***Nowadays simulations are far from the solar parameter regimes*** (e.g., [9]). And as the resolution is increased, parameters are generally not held constant. Even worse, a fundamental question is unanswered: How much resolution is enough? In particular, higher resolution allows for higher Reynolds, magnetic Reynolds ($R_m$), and Peclet numbers, and as these parameters are increased, the flow generally becomes more turbulent and the convective patterns and transport properties may change.

Also, a goal for solar convection simulations is to resolve all scales which are significantly influenced by rotation and stratification (i.e., buoyancy) in order to capture all the self-organization processes. In addition, the transition regions which couple the convection zone to the radiative interior below and the solar atmosphere above are particularly challenging to resolve in global simulations. Solar convection simulations must always push the limits of available high-performance computing platforms to achieve ever higher spatial resolution. However, the highest-resolution simulations achievable on a given platform are computationally intensive. Thus, it is impractical to run the highest-resolution simulations for the long durations necessary to adequately assess sustained dynamo action or to explore dynamics spanning several solar activity cycles. For such investigations, intermediate-resolution simulations will always remain important. And the latter further emphasizes the need for reliable subgrid-scale models to account for motions which are not resolved at present. Nowadays, one can look at very good 3D numerical MHD experiments studying the dynamo action in turbulent flows, as for instance the work of Archontis et al. [17], where the ABC flows were adopted as the candidate velocity field. However, knowledge of the exact mechanisms behind the maintenance of magnetic fields against resistive diffusion is still incomplete.

Finally, even though this brief introduction may convince anybody that beyond any doubt the dynamo of the Sun will continue to be an enigmatic issue; it is astonishing that a massive celestial body is almost perfectly organized producing a "clock-precision" periodicity. The Sun during its lifetime (of about $10^{10}$ years) will actualise $\sim 10^9$ sunspot-cycles; a number that approximately equals the total heartbeats of a man of 30 years old. And such a response certainly requires a very unusual mechanism. In any case, the better understanding of the involved physical processes is always a priority. As we think, there is presently an arbitrary hypothesis related to the following big and unanswered question: How can an astrophysical object such as the Sun generate a systematic (large-scale) magnetic field at high $R_m$? How can it overcome its tendency to be dominated by fluctuations at the small scales? We think that the fast dynamo theory requires something more than the assumed convection plasma motions, which may have been historically influenced by theories concerning the Earth's dynamo. The overemphasized "velocity field" along with the current MHD approach potentially leading to the fast dynamo action may be actually an arbitrary assumption. In the proposed dynamo model the key role is played not by convection; in contrast, as we shall show in this work, the two toroidal currents sustained inside the Sun's tachocline and flowing with zero resistivity essentially modulate the whole 22-year solar cycle.

# 2. A few fundamental processes

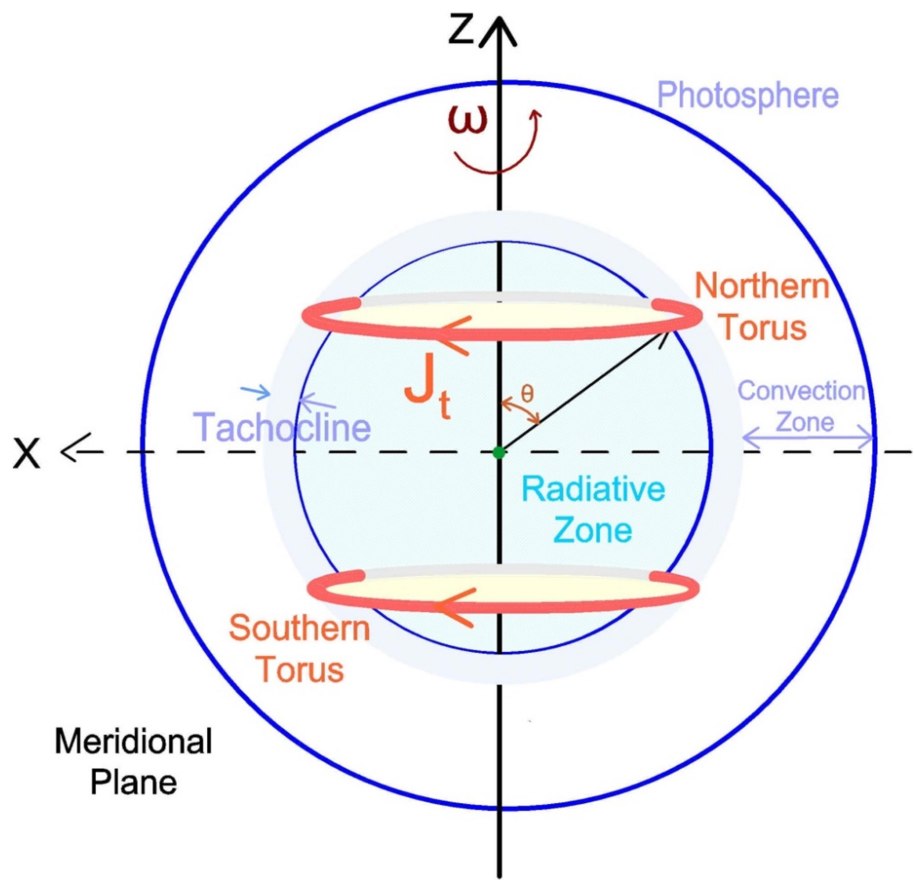

**Fig. 1.** Schematic showing a solar meridional plane with the Radiative Zone (RZ), the tachocline, the Convection Zone, and the two toroidal structures (inside the tachocline) with westward azimuthal currents.

In this work, we suggest the fundamental processes potentially establishing the mechanism of "fast dynamo action" in the Sun. The "key entity" with outstanding, extraordinary and unique properties, in the Sun's interior, is a complex of two tori formed within the tachocline: One Torus for each hemisphere symmetrically placed with respect to the equatorial plane (**Fig. 1**). And each Torus, made up of toroidal and poloidal magnetic fields, has three prominent properties: First, it remains in a stable state for almost the whole 11-year sunspot cycle; second, it carries a surplus charge that changes sign in the next sunspot cycle and, third, it is characterized by huge amounts of currents flowing with zero resistivity. The outcome reveals a "fast dynamo" action, based on the Sun's ability to build up, in its interior, a charged structure for such a prolonged time period. The two toroidal structures are generated in the tachocline composed of completely ionized and very high density plasma. Most importantly, we show in a convincing manner, that these two structures provide the possibility for a reverse of the magnetic polarity, within the 22-year solar cycle. In the following paragraphs, we progressively approach several of the key aspects of the fast dynamo model; it is unavoidable that (during the first reading) open issues will remain until the whole concept-scenario is presented and all the basic processes are scrutinized. Then, we shall go on synthesizing a simplified "3D solar circuit".

## 2.1. The charged sphere of the Radiative Zone (RZ)

We initially suppose a northward directed "seed magnetic field", $B_{zo}$, hypothetically being the galactic magnetic field in which the Sun is immersed. This field does not penetrate into the Radiative Zone (RZ) assumed being a perfect conductor (**Fig. 2**). As a matter of fact, surface currents azimuthally flowing probably force the magnetic field lines to envelop the RZ. Only over the RZ-polar regions, the magnetic field is roughly directed radially outward. When the RZ is under charging (via a specific mechanism exhibited below), the growth of a uniform surface charge density would be the anticipated response. In the latter case, of a uniformly charged surface having charge density $\rho$ (cm$^{-3}$), radius $r_o$ and rotating with the solar angular frequency $\omega$ (rads$^{-1}$), the produced current density $J_{\varphi}(\theta)$ in a surface layer will be

$$J_{\varphi}(r, \theta) = \omega \rho r_o \sin\theta.$$

And the resulting magnetic field outside the RZ will be pure dipole (due to the specific relation between $\sin\theta$ and $J_{\varphi}$). The perpendicular to the RZ-surface electric field will be $E = \rho/\varepsilon$ (Vm$^{-1}$); as it is shown in **Fig. 2** for a negatively charged RZ. In contrast, outside the RZ, the resistivity will monotonically increase with the distance, and out of the photosphere it will become extremely high. The particle number density of the photosphere is $n_{photo} = 10^{23}$ m$^{-3}$, while the plasma is partially (~3%) ionized. The number density at the tachocline is much higher; $n_{tacho} = 10^6 \, n_{photo}$, while the plasma is completely ionized.

Consequently, over the polar regions of the RZ, *the electric field is roughly in parallel to the magnetic field, while any magnetic force vanishes* since the solar rotational speed is locally almost zero. Thus the electrons can move radially outwards from the (negatively charged) RZ along a "conductive channel". And the channel remains open, since an inward directed gradient for the electron number density is established and associated with intense electrostatic repulsive forces. From the rest of the sphere's surface any outward electron diffusion is impossible, given that the magnetic force $\mathbf{F_L}=-e\mathbf{u_\varphi}\times\mathbf{B_{zo}}$ (outside the flanks of the RZ-sphere) is inward directed; where $u_\varphi$ is the (azimuthal) rotational speed of the Sun for each electron.

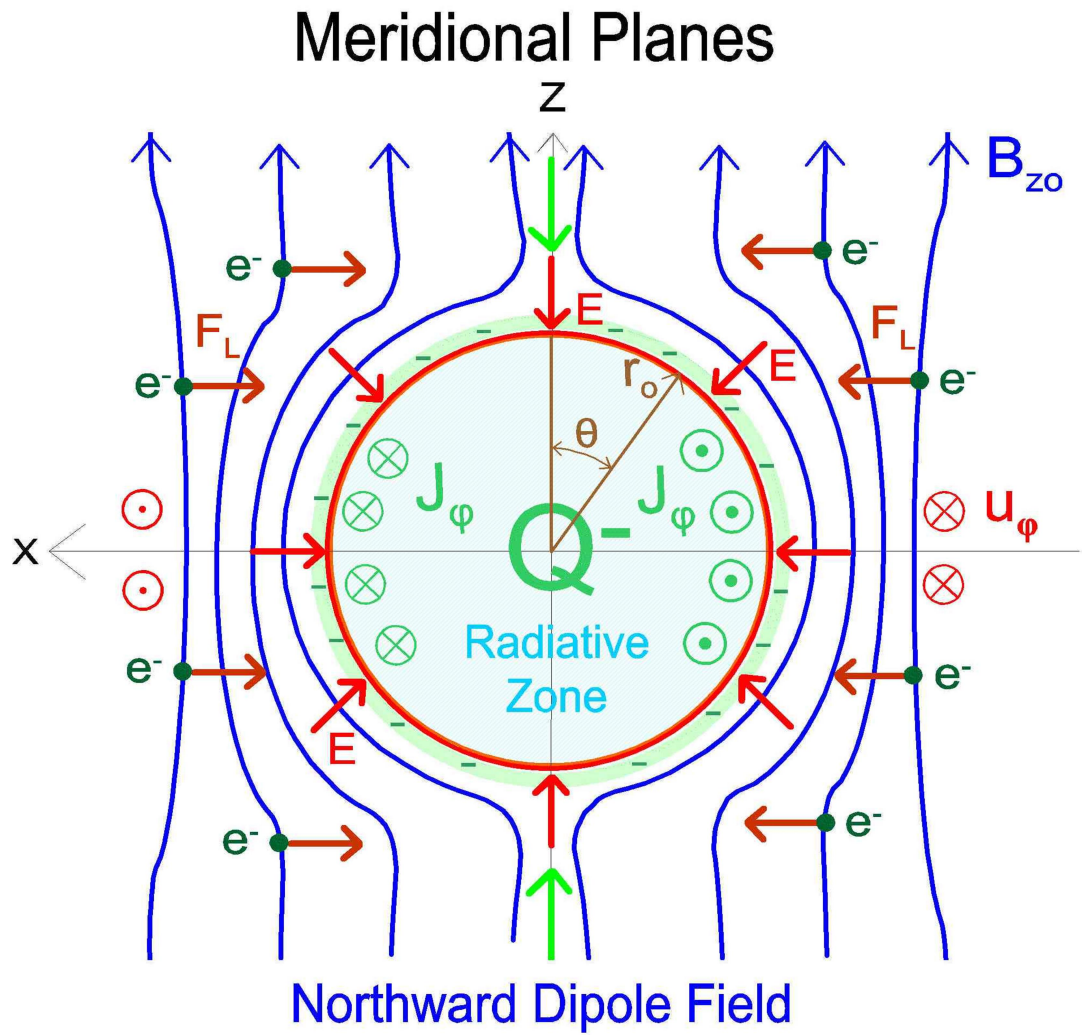

**Fig. 2.** The seed magnetic field, $B_{zo}$, enveloping the perfect RZ-conductor, forces the electrons via the Lorentz–magnetic force to move inwards. Thus the sphere of the RZ is negatively charged. The uniform surface charge density produces, over the polar regions, a radially inward directed electric field, which is roughly in parallel to the local magnetic field. Electrons can escape from the poles, since there is an intense (inward directed) gradient for the local electron number density.

## 2.2. Why the RZ is negatively charged (under northward directed dipole field)

Each electron within the Convection Zone is subject to the Lorentz-magnetic force, $\mathbf{F}_L=-e\mathbf{u}_\varphi \times \mathbf{B}_{zo}$, due to the seed magnetic field, $B_{zo}$, (which is arbitrarily chosen being northward directed) and moves inward (**Fig. 2**). In effect, the RZ is negatively charged. The potential difference (at low latitudes) between the photosphere and the RZ will be $\Delta V=E\ \Delta L=u_e B_{zo}\Delta L$, where $u_\varphi$ is the rotational speed, and $\Delta L=0.3\ R_{sun}$. Hence, a gigantic capacitor is built up with the positive charge dwelling in a layer adjacent to the photosphere and the negative plate essentially being the surface of the RZ (**Fig. 3**). Eventually, the Sun's rotation beside the involved charge separation process will obviously reinforce the initial seed magnetic field. The $B_{zo}$ increases to $B_{z1}$ and given that more electrons move inwards, the $B_{z1}$ will further increase. In this way, the process occurs in a repetitive fashion. Therefore, a "slow dynamo" action results, since the resistivity within the Convection Zone is far from being zero. If we assume a polar magnetic field $B_{z1}=1\ G=10^{-4}\ T$, and an average electron-rotational speed $u_\varphi=1700\ ms^{-1}$, then the potential difference across the Convection Zone will be $\Delta V=35\ MV$.

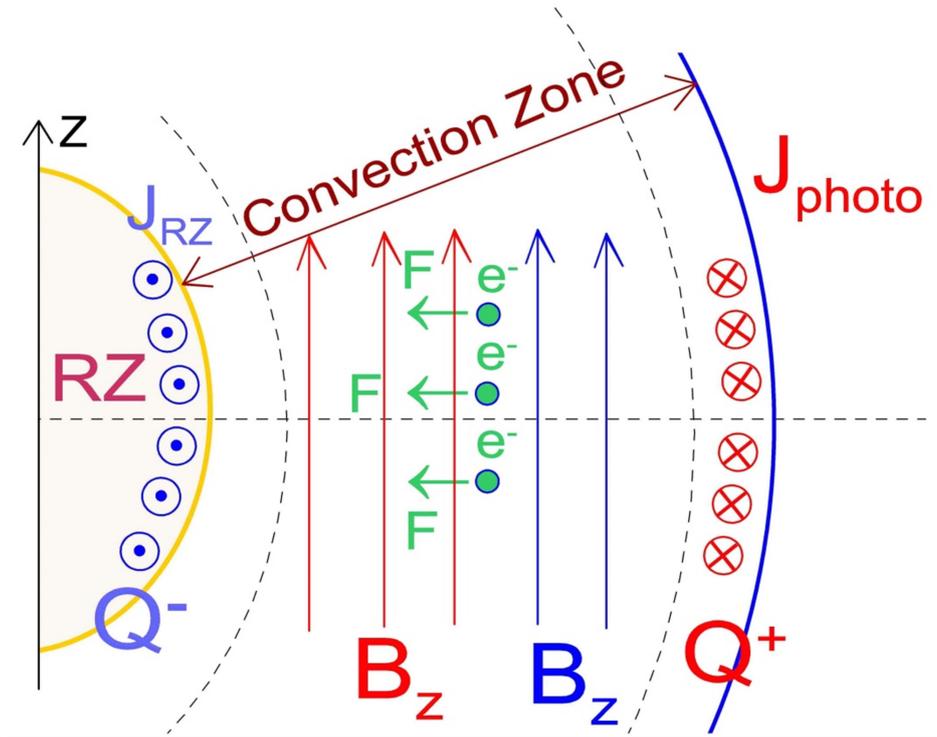

**Fig. 3.** Each electron within the Convection Zone is subject to the Lorentz force, $F_L=-eu_\varphi \times B_{zo}$, due to a seed northward directed magnetic field, $B_{zo}$, and moves inwards; $u_\varphi$ is the rotational speed. In effect, the RZ is negatively charged. Hence, a gigantic capacitor is built up with the positive charge dwelling in a layer adjacent to the photosphere and the negative plate essentially being the surface of the RZ. The $B_{zo}$ increases to $B_z$, a process occurring in a repetitive fashion.

In general, the helioseismic and surface observations of the solar differential rotation present several compelling challenges (e.g., [9]); we briefly mention here two of them, while one more is added in the discussion subsection 4.3.

(a). Throughout the convective envelope, the rotation rate decreases monotonically toward the poles by about 30%. Angular velocity contours at mid-latitudes are nearly radial.

(b). Near the base of the Convection Zone, there is a sharp transition between differential rotation in the convective envelope and nearly uniform rotation in the radiative interior. This transition region has become known as the solar tachocline wherein the Torus (actually the two toroidal structures) is probably formed. The rotation rate of the radiative interior is intermediate between the equatorial and polar regions of the Convection Zone. Thus, the radial angular velocity gradient across the tachocline is positive at low latitudes and negative at high latitudes, crossing zero at a latitude of about 35°.

These features play a substantial role in the presented model demonstrating why the $\Delta V$ is essentially developed at lower latitudes, between the RZ and photosphere. That is, at latitudes corresponding to the sunspot area, which is limited to ±35°. Moreover, a negative charge accumulated in the tachocline (and due to the ω-shear) may clearer demonstrate how a huge capacitor is built in the Sun's interior.

## *2.3. About the Torus formation*

As the electrons escape radially from the poles of the RZ, a circular magnetic field, according to Ampere's law, is developed around this electron current. In this way, an early toroidal magnetic field is formed outside the RZ and close to the north and south poles. Eventually, this "seed toroidal field" is evolved building up the two stable toroidal structures. As "Torus" (i.e., a single-holed "ring") we mean a surface of revolution generated by revolving a circle (of radius $R_{torus}$) in three-dimensional space about the Sun's (north-south) Z-axis coplanar with the circle. Initially, we envision a toroidal magnetic field with small radius placed at a specific place (outside the RZ), in which "the stability condition" is satisfied. And the stability condition is extensively studied bellow. The electrons penetrating into the Torus (and having the rotational speed, $u_\varphi$, of the Sun), automatically, set up an azimuthal current as they are "trapped inside". Each electron, of the surplus Torus charge, constitutes an elementary toroidal current and two electrons, although being close together, do not repel each other; and the latter is dictated by the stability condition.

Once a "seed toroidal current" is formed, it rapidly grows. Torus reaches its mature phase while drifting to lower latitudes, and finally disintegrates close to the equatorial plane. However, all these issues must be thoroughly studied, and we shall inevitably re-examine the Torus generation mechanism in a dedicated subsection later on. Right away, we go ahead presenting the stability condition of electrons inside each Torus, as this claim is the most prominent of the work.

## *2.4. The stability condition for a negatively charged Torus*

The fundamental theory of electrodynamics defines that the speed of an electromagnetic disturbance, in a medium with permeability μ and permittivity ε, is

$$c' = (\mu\varepsilon)^{\frac{1}{2}}$$

In a vacuum, $c'$, is the "free space" speed of light; $c'=c=3*10^8$ ms$^{-1}$. The Sun's toroidal structures are presumably developed at those well-defined places where ***the speed of electrons equals the local speed of an electromagnetic wave***; *that is, the electrons of the surplus charge satisfy the condition*

$$u_e = c'$$

We suppose that the Torus is formed in the tachocline (i.e., very close to the Radiative Zone), at a distance $\sim 0.7$ $R_{Sun}$ from the Sun's center. Over the surface of the RZ, $c'$ is almost zero, whereas the $u_e$ is essentially the speed due to the solar rotational speed (i.e., $u_e=u_\varphi$ throughout the work). Over the photosphere, $c'=c$.

From a stationary reference frame, an electromagnetic disturbance within the Torus will move eastward with velocity $2c'$, given that the local rotational speed is $u_e=u_\varphi=c'$. In this situation obviously the westward velocity is zero. The electromagnetic wave velocity for an observer positioned in the Torus plasma and moving with velocity $u_\varphi$ will be $c'$ for all the directions. For both frames, the electromagnetic velocity perpendicular to the rotational speed $u_\varphi$ will always be $c'$. Therefore, ***having two electrons (with the same longitude) within the Torus, the force of interaction is zero, when $u_e= c'$; that is, the electrostatic force counterbalances the magnetic force***. In this situation the ratio of electrostatic to magnetic force is as follows:

$$F_e/F_m = (e^2/4\pi\varepsilon r^2)/(\mu e^2 u_e^2/4\pi r^2) = 1/(\mu\varepsilon u_e^2) = (c'/u_e)^2$$

Consequently, we can accumulate more and more electrons within the space of Torus. The (repulsive) electrostatic force exerted on a new inflowing electron by the Torus overall charge equals the (attractive) magnetic force caused by the Torus current on the same test charge. Practically, we would assume the Torus as an infinite line with a uniform charge density or as a 3D space with a volume charge density. However, the Torus structure is more complicated and the next paragraph emphasizes a few critical aspects. For the next sunspot cycle, the same principle should be applied for ***an electron-deficient core Torus region***. We usually deal with a negatively charged Torus; nevertheless, the fundamental behaviour remains the same for a positively charged Torus.

## *2.5. The Torus internal structure*

We consider a negatively charged Torus drifting to lower latitudes, while remaining on the trajectory determined by the relation $u_e=c'$. In this way, the Torus is flanked by two well-defined regions; the first one positioned closer to the RZ and characterized by the relation $u_e>c'$ and the second one with $u_e< c'$ (**Fig. 4**). In the former region the magnetic force prevails, whereas in the latter one the electrostatic force is the dominant one. Moreover, we would envisage the Torus as composed of the regions R1 and R2, plus a homogenously overcharged "core region", R0. The latter, ***although is negatively overcharged, none electron is repelled outwards***. Assuming that all the electrons of R0 obey in the relation by $u_e=c'$, then any supposed repulsive electric force is counterbalanced by an attractive magnetic force. All the Torus electrons, having roughly the same azimuthal speed $u_e=u_\varphi$, would be considered as ***elementary homoparallel currents subject to attractive magnetic forces***. The overall westward electron current of Torus produces a strong poloidal magnetic field dominating in the regions R1 and R2, and producing an attractive force. We shall discuss latter on, if R0 is potentially "an electron-degenerate region". An asymmetry existing between the regions R1 and R2 is considered in the next subsections. Any possible estimate concerning the Torus (cross-sectional) radius or the toroidal current density is beyond the scope of this work.

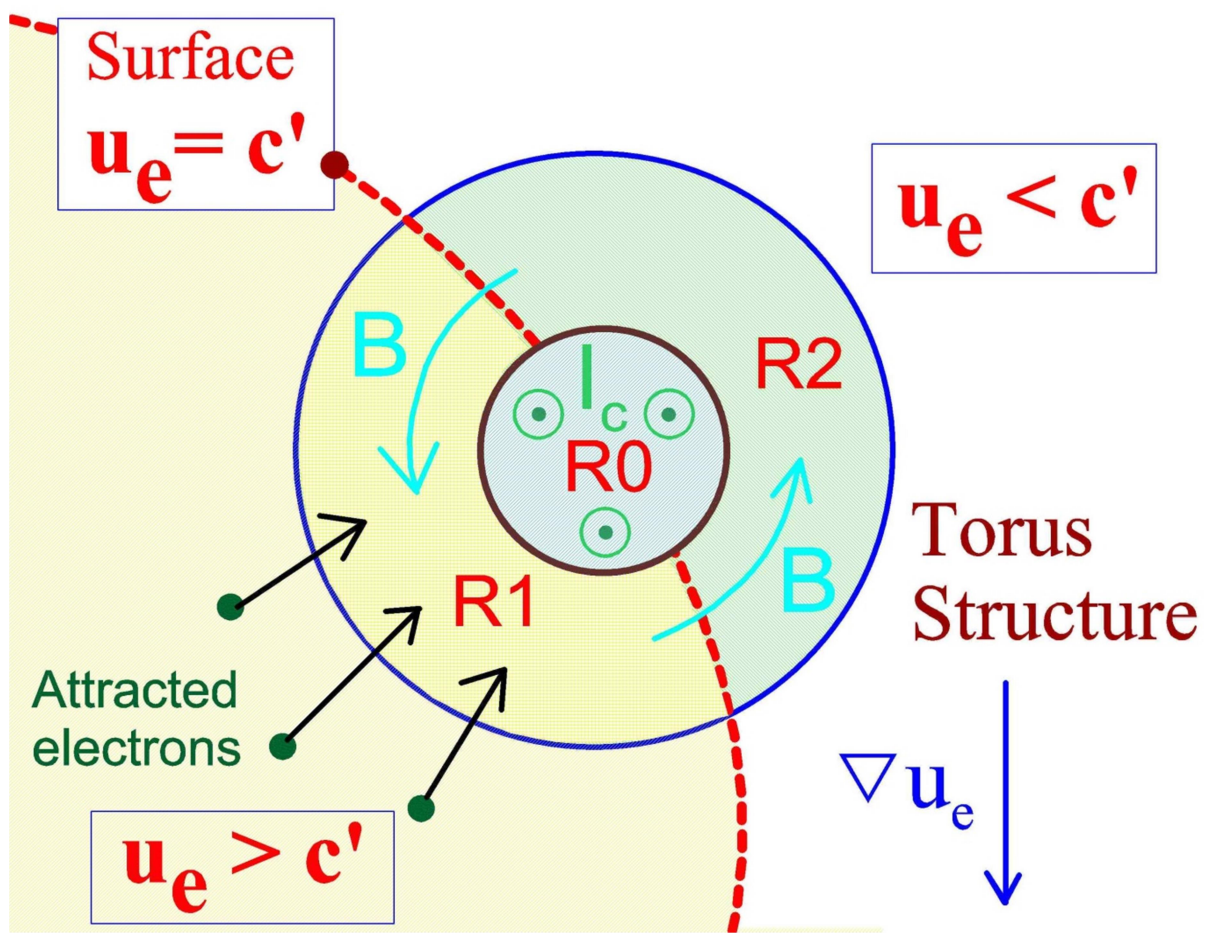

**Fig. 4.** A negatively charged Torus is composed of the regions R0, R1 and R2. In the region R0, an intense azimuthal-westward current ($I_c$) flows, producing a strong poloidal magnetic field in the regions R1 and R2. The electrons in the Torus-core region, R0, obey in the relation $u_e = c'$; that is, R0 is the region wherein the electrostatic repulsive force equals the attractive magnetic force. The regions R1 and R2 are separated by the surface $u_e = c'$ (being the geometric place on which the electron azimuthal speed equals the electromagnetic wave speed). This "trajectory" also dictates the orbit of Torus drifting to lower latitudes with speed ~2.5 ms$^{-1}$. Electrons supplied to the R0 region are essentially pumped from the region R1 and R2 via the poloidal magnetic field.

## 2.6. The "Debye anti-shielding" effect

Plasmas generally do not contain strong electric fields in their rest frames. In astrophysical plasmas, Debye screening prevents electric fields from directly affecting the plasma over large distances, i.e., greater than the Debye length. The shielding of an external electric field from the interior of a plasma can be viewed as a result of the polarization of the plasma medium and the associated redistribution of space charge, which prevents penetration by an external electric field. The length-scale associated with such shielding is the Debye length. However, this consideration implicitly assumes that there are many particles in the shielding cloud. And the latter is impossible (as we shall exhibit below) for the two toroidal structures of the Sun and the region very close to the RZ. Moreover, the existence of moving charged particles causes the plasma to generate, and be affected by, magnetic fields, and this can and does cause extremely complex behaviour. Also we know that it is possible to produce a plasma that is not quasineutral. An electron beam, for example, has only negative charges. However, the density of a non-neutral plasma must generally be very low, or it must be very small, otherwise it will be dissipated by the repulsive electrostatic force.

The answer to all these issues, concerning both the RZ and the two toroidal structures, is studied in detail below. We have previously supposed that the RZ is negatively charged (and the situation is essentially not different with a positively charged RZ, occurring at the next sunspot cycle), while remaining immersed in dense and completely ionized plasma. ***Any "shielding" of the charged RZ is prevented*** and the reason is the following: First, the net surface charges of the RZ produce surface azimuthal currents (due to the solar rotational speed), and a northward directed magnetic field skimming the surface of the RZ (**Fig. 5**). Second, at least at lower latitudes of tachocline (that is, in a zone extended to $\pm 35^{\circ}$) the magnetic force dominates over the electrostatic one, given that this region is clearly placed inward of the curve $u_e = c'$. As a matter of fact, the relation $u_e = c'$ defines an ellipsoid of revolution (red-dashed line); inside (outside) the spheroid's surface dominates the magnetic (electric) force. That is, ***an electron placed inside the spheroid is magnetically attracted inward; and eventually it is added to the negative charge (Q') of the RZ***. However, there is no any possibility that a "double layer" might be generated between the RZ and the Convection Zone. Always the electrons will flow inward further increasing the charge of the RZ. Any shielding process is absolutely prevented.

In the Torus situation, the "inner region" termed R0 is assumed as "negatively overcharged", although existing within the ionized plasma of tachocline. If positive charges were hypothetically accumulated around the region R0, then the entry of new electrons would have been prevented. Consequently, the electrons inside the Torus would be screened out; the toroidal current would not grow and the associated poloidal magnetic field would become finally zero. However, this is not the situation. We remind again that the Torus is crossed by the spheroid defined through the relation $u_e = c'$; thus, over the one Torus flank (which is placed adjacent to the RZ), the dominant magnetic force absolutely prevents any shielding much like to the situation around the RZ.

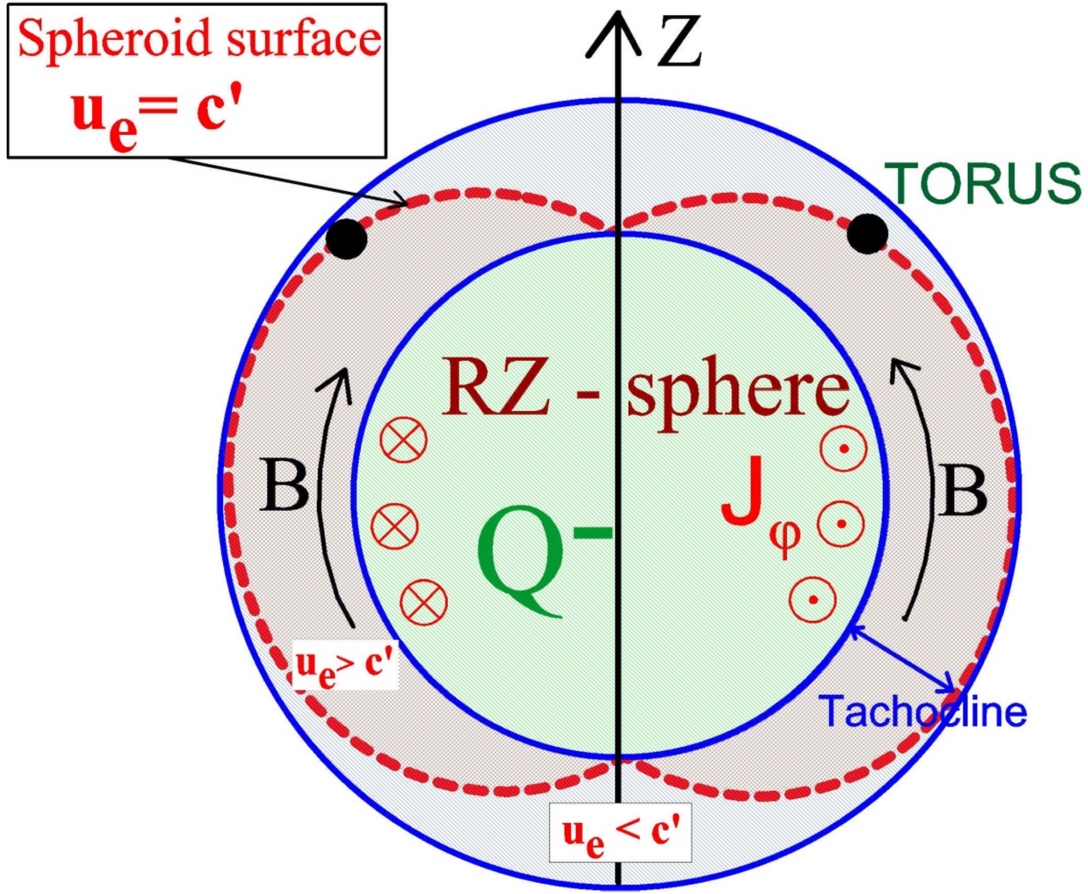

**Fig. 5.** A meridional cross section of the RZ-tachocline complex showing the northern Torus (as black-solid circles) drifting equatorward along the red-dashed trajectory defined by the relation $u_e=c'$. The (negatively charged) Torus remains within the tachocline (being the area between the two blue circular lines) that envelops the RZ (green shaded area). The blue (and blue-red) shaded area of tachocline denotes the region with $u_e<c'$ (and $u_e>c'$), wherein the electrostatic (the magnetic) force prevails. The azimuthal current $J_\varphi$ flows westwards on the RZ-surface, given that the RZ is negatively charged (in the case of northward directed dipole magnetic field).

It may be useful to visualize the three different plasma reactions occurring within the (negatively charged) Torus and adjacent of the RZ plasma regime. **Figure 6** schematically shows the three plasma categories with a translational velocity $u_e$, being perpendicular to the paper-plane, and the velocity $c'$ of an electromagnetic wave. The placement of a negative charge $Q^-$, within each plasma regime, causes ***(a) Debye shielding ($u_e<c'$), (b) "Debye anti-shielding" ($u_e>c'$) and (c) the particle distribution remains unaffected ($u_e=c'$).*** The "anti-shielding" effect charges the surface of the RZ, as well as the two toroidal structures. The reader must assimilate the idea that apparently there is no any shielding for a charge placed within the Torus-core region, since $u_e=c'$. In this situation, the electrostatic force (attractive or repulsive) equals the magnetic force (repulsive or attractive, respectively).

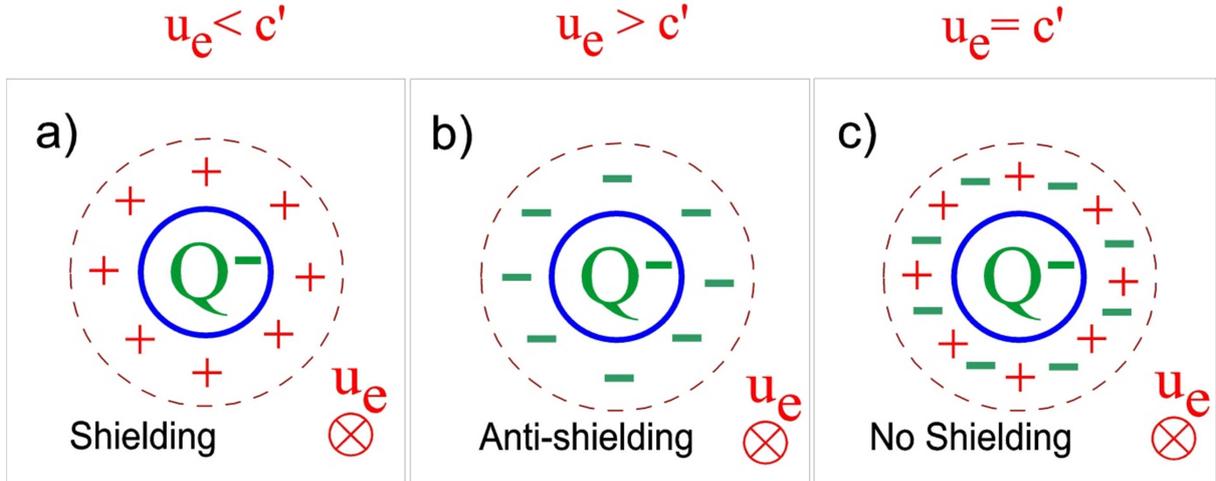

**Fig. 6.** Three categories of plasmas (with $u_e<c'$, $u_e>c'$ and $u_e=c$) are subject to the same translational motion with velocity $u_e$ (perpendicular to the paper-plane). The placement of a negative charge $Q^-$, within each plasma regime having the same velocity $u_e$, causes (a) Debye shielding ($u_e<c'$), (b) "Debye anti-shielding" ($u_e>c'$) and (c) an unvaried particle distribution ($u_e=c'$).

One has to pay attention that the RZ is homogeneously charged despite the validity of relation $u_e>c'$. If the magnetic forces increase the charge density locally, then the same forces will counteract reducing the density.

## 2.7. The Torus equatorward drift

We consider again a negatively charged Torus; the increase of the toroidal current is associated with an additional process. In the neighbourhood of each Torus, the magnetic force is greater at lower latitudes, given that there is an equatorward gradient for the azimuthal electron velocity $u_\varphi$ due to the solar differential rotation. In addition, the magnetic force prevails against the electrostatic one, only in the region R1 of **Fig. 4**. Thus, electrons essentially flow into the Torus core region, R0, exclusively from the lower latitudes of region R1; an asymmetry that may account for the equatorward Torus drift. Moreover, one may argue that the tori are mutually attracted because of their homoparallel currents. At the end of an 11-year cycle, the two tori arrive at the equatorial plane and release their net negative charge, while their oppositely directed magnetic field lines disintegrate. In conclusion, throughout the Torus equatorward drift, a surplus of negative charge is trapped in both toroidal structures. The Torus drift velocity, lasting ~10 years, is ~2.5 ms$^{-1}$.

## 2.8. The Torus generation

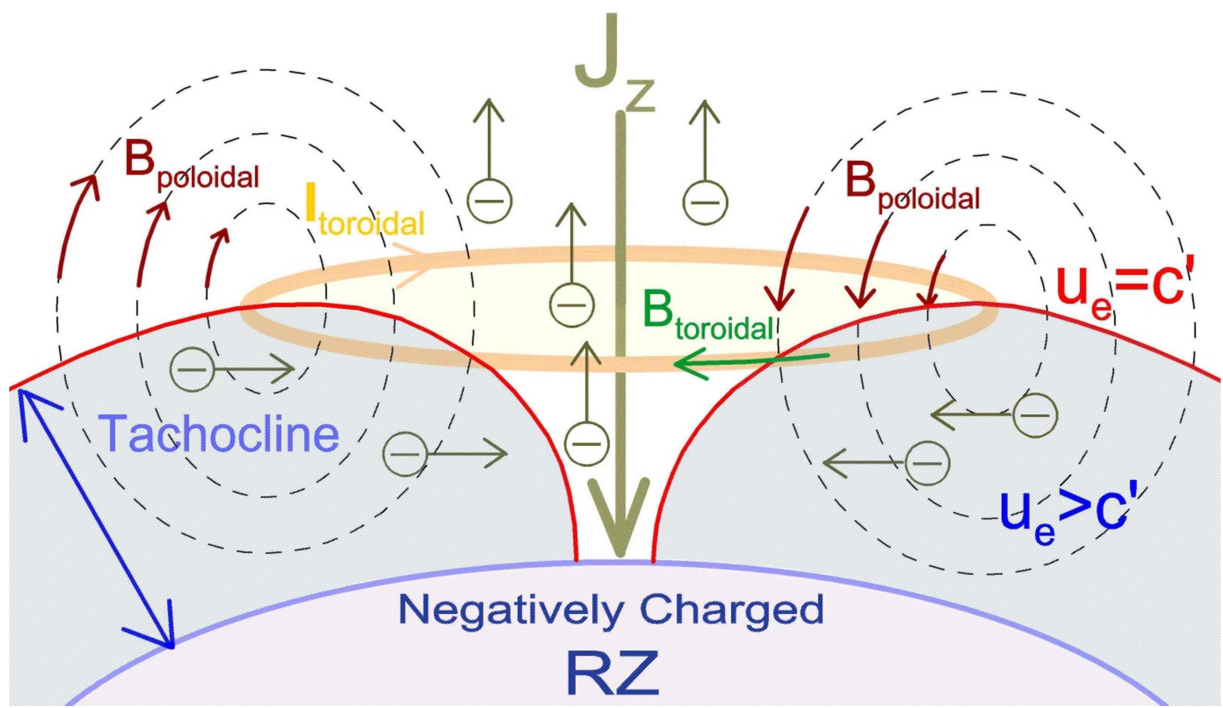

**Fig. 7.** The early (or seed) toroidal magnetic field (B$_{toroidal}$) is probably formed within the tachocline (from the radial current J$_z$ of the escaping electrons) and over the surface where the condition u$_e$=c′ is satisfied. The toroidal current produces the large scale poloidal magnetic field that attracts additional electrons from the region where u$_e$>c′ (blue-shaded area). Thus, the Torus current rapidly grows to very high intensities.

The negatively (positively) charged RZ are essentially associated with negatively (positively) charged Tori. Presently, for a negatively charged RZ, electrons from the RZ easily escape from the polar areas radially outwards, producing an "exodus channel". Especially, they rapidly move within the conical region wherein the electrostatic force dominates against the magnetic one (u$_e$<c′, **Fig. 7**). In addition, in the polar region the magnetic field is almost in parallel to the electric field being perpendicular to the RZ. Thus, the escaping electrons form a current J$_z$ flowing along the solar rotational Z-axis and producing the necessary toroidal magnetic field which is a significant ingredient of the Torus structure. The toroidal magnetic field is maximized just outside the electron channel. The tachocline electrons, around the north pole region confined between the surface $u_e=c'$ and the RZ, move perpendicularly to the Z-axis and finally charge the toroidal magnetic field structure. Electrons probably penetrate into the embryonic Torus and remain trapped; the Torus moves over the spheroid surface defined by the relation $u_e=c'$.

The Torus, with its own toroidal current, produces the required poloidal magnetic field that strongly attracts additional electrons from the region where $u_e>c'$. Each electron, outside the Torus, is attracted by the much higher population of toroidal electrons; ***the electrons inflow like an avalanche***. Thus, the Torus current rapidly grows and reaches its mature stage. The attracted electron population is higher from the portion of the Torus placed further away from the Z-axis. In the case of southward directed magnetic field, the electrons produce an electron-deficient "embryonic Torus" and the poloidal magnetic field continuously subtract electrons outside the (already positively charged) Torus.

## 2.9. *Huge toroidal currents formed in the Sun's interior*

In the case of the Sun, both of the tori are created symmetrically with respect to its equatorial plane (**Fig. 1**). The distance R of each Torus from the rotational axis is varied given that, as the time passes, each Torus moves equatorwards. The tori are initially formed close to the poles of the RZ; then steadily move and finally disintegrate close to the equatorial plane, at the end of each 11-year sunspot cycle. They always create oppositely directed toroidal magnetic fields within the same solar cycle; and this is dictated by the radially (outward or inward) directed electron flows along the Z-axis in the Convection Zone. In each Torus the poloidal magnetic field is produced from the toroidal current (westward or eastward when the dipole magnetic field is directed northward or southward, respectively). As the Torus charge grows like an avalanche, an intense toroidal current is established, ***since the Torus charge is revolving with the Sun's rotational speed***. Obviously, as the R increases and the Torus electron velocity increases, too. For instance, if R=0.74 $R_{sun}$ (given that the radius of the RZ is 0.7 $R_{sun}$ and the thickness of the tachocline is 0.04 $R_{sum}$), then the maximum electron speed (over the equatorial plane) will be $u_\varphi \approx 1500$ ms$^{-1}$. If R=0.4 $R_{sun}$ (i.e., the Torus is obviously off the equatorial plane), then $u_\varphi \approx 800$ ms$^{-1}$. Given that the $u_\varphi$ is many orders of magnitude greater than the "drift electron velocity" associated with the typical metallic conductors on Earth, we infer that the Sun's toroidal currents reach huge values.

In both of the toroidal structures, that is, the Torus structure in tokamaks and the solar Torus structure, toroidal and poloidal magnetic fields and currents are observed alike.

## 2.10. The electron "confinement time" within a negatively charged Torus

The mean free path, $\lambda_c$, of electrons due to Coulomb scattering for the tachocline (and the photosphere as well) is very small. In tachocline, if the electron number density is $\sim 10^{23}$ cm$^{-3}$ and the temperature T=$10^6$K, then $\lambda_c$=2.6 nm. In the photosphere, with an ionization percentage of 3 %, n=$3*10^{15}$ cm$^{-3}$ and T=5000 K, the $\lambda_c$ is 2 μm. Thus, once an electron gains its entry into the Torus, then it will remain "trapped for ever" having the local rotational speed of the Sun. In terms of the "confinement time", in a tokamak plasma reactor, the situation of the Sun with a confinement time of about 11 years is beyond any currently mental image. In tokamaks, we envision a confinement time of 1000 s as a future goal, for instance, in the case of the "International thermonuclear Experimental Reactor", ITER),

## 2.11. The zero resistivity of toroidal currents

The major challenge, the paradox taking place in the Sun's interior, is that **two tori are developed and remain charged for ~11 years**. Moreover, the rotational speed of the Sun, being in this extraordinary case the electron velocity, is absolutely independent from the matter itself. Thus, there is no any resistivity, **and the resulting dynamo effect is really a fast dynamo one**. In this perspective, the thought-provoking question is what actually happens with the repulsive electron forces, for instance, inside each negatively charged Torus. Everyone anticipates that the strong repulsive electrostatic forces will immediately destroy any local excess of net charge. Could we envision a stable Torus structure charged with trillions of Coulombs for such a prolonged time period? And the answer is already given in the preceded subsections.

## *2.12. The Torus disintegration*

The long-lived stability condition within the Torus (i.e., $u_c = c'$) gradually becomes susceptible to impairment and finally breaks down. The latter is obviously accomplished through a specific mechanism, on which we draw our attention.

Let us suppose that the Convection Zone magnetic field is northward directed and the two tori drift equatorwards. At the same time, negative charge leaks out from the poles of the RZ; and finally there is a moment at which the total magnetic field close to the RZ becomes southward directed. The latter essentially is resulted from the total poloidal magnetic field of the two (negatively charged) tori. In this way, the RZ is switched to positive charging. When the tori collide over the equatorial plane and form just one new Torus with double negative charge, one may think that the newly formed Torus would continue to fulfil the same stability condition and probably survive. In contrast, the new toroidal structure rapidly disintegrates and loses its charge; the electrons move outwards and disperse widely. The main reason of disintegration is the annihilation of the toroidal magnetic field in the new Torus; given that the individual toroidal fields are always oppositely directed. It should be underlined that the toroidal magnetic field is probably a stability factor for the "core region" termed "R0" in subsection 2.5. Every electron having a radial thermal velocity $u_p$ will be subject to the force $e\mathbf{u_p}x\mathbf{B_{tor}}$ that runs perpendicularly to the velocity. The latter obviously reduces the outward electron diffusion from the region R0. Thus, the toroidal field does not allow the exit of electrons out of the Torus. The antiparallel currents developed in the RZ and the new Torus, on the equatorial plane, are mutually repulsive (**Fig. 8**). In our schematic, the Torus is formed within the outer tachocline. If the tachocline has a thickness of 0.04 $R_{sun}= 4.3$ $R_E$, then we could envision that the Torus diameter may be 1-2 $R_E \approx 10$ Mkm. The Torus is immersed in a southward ($-B_z$) directed magnetic field forcing the "trapped electrons" of the core region to move outwards. These electrons violate their stability condition and the resulting electrostatic forces (within the Torus region R2) damage (beyond any repair) the Torus entity. After the Torus destruction, the only remaining entity is the positively charged sphere of the RZ surrounded by the tachocline. Then, the electrons of the Convection Zone move farther outwards affected by the locally southward directed magnetic field; consequently, the positively charged region constantly widens. Eventually, the positively charged sphere becomes the widest one covering almost totally the Convection Zone. The electrons are displaced in a sub-surface thin layer (beneath the photosphere) creating a westward directed electron current. Certainly, the just described scenario is a first approach of the Torus disintegration process; the latter probably begins earlier than the two tori "collision time".

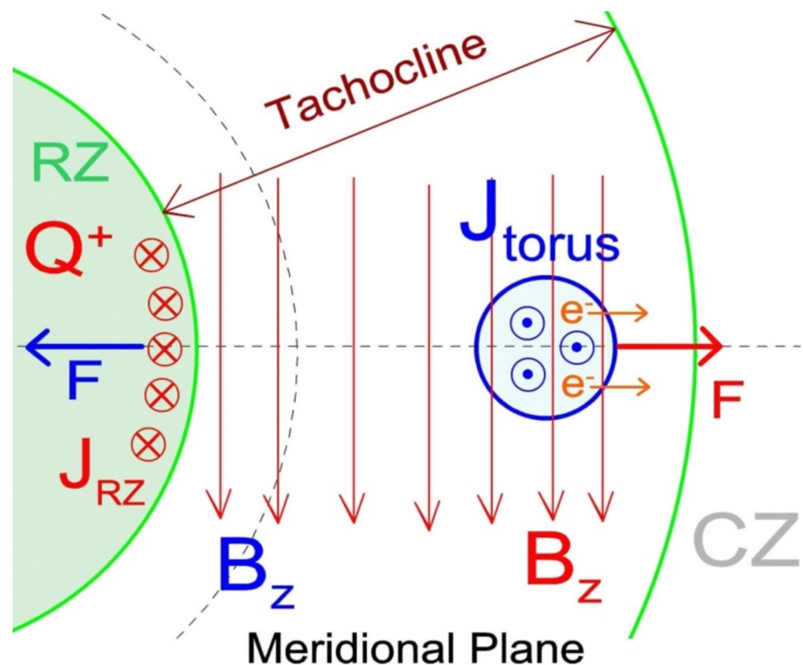

**Fig. 8.** During the equatorward Torus drift, the initially attractive magnetic force between the negatively charged RZ and the Torus gradually weakens. When the RZ is switched to positive charging (as in this sketch), this force becomes repulsive. After the tori collision, the "new Torus" is immersed in a southward (i.e., $-B_z$) directed magnetic field, and the "trapped electron population" inside the Torus is forced to move outwards. The latter violates their stability condition; the electrons go through the Torus region R2, wherein the prevailing electrostatic forces damage the very Torus structure.

## 2.13. The polarity of the dipole field

Essentially the dipole magnetic field is determined by the sign of charge of the RZ. The negatively (positively) charged RZ is associated with northward (southward) directed dipole magnetic field. The latter will become further clear in the next section, where all the successive processes are presented for a typical 22-year solar cycle. The static charge of the RZ produces a surface azimuthal current due to the Sun's rotational speed. In general, the magnetic field within the Convection Zone results as the superposition of fields produced by all the three azimuthal currents: (a) the $J_{RZ}$ from the surface charge of the RZ, (b) the $J_{photo}$ from the charge accumulated in the layer beneath the photosphere, and (c) the two toroidal currents of the charged toroidal structures.

## *2.14. The Torus equatorward drift mechanism in agreement with the Waldmeier rule*

The mechanism through which the Torus drifts equatorward (subsection 2.6) easily explains the Waldmeier rule [18]. Given that any single cycle is asymmetric in time, then according to the "Waldmeier rule" the strong cycles are faster to reach their activity maximum than the weak cycles. And this rule is used to produce successful predictions of the strength and duration of future activity cycles.

The explanation, in the context of this work, is that as the Torus current increases, the mechanism attracting or repealing electrons inside the "core region", R0, becomes more efficient (i.e., faster) through the Torus poloidal magnetic field. Finally, the higher rates for the inflowing or outflowing electrons imply higher values for the (equatorward) drift velocities of the Torus itself (subsection 2.6). In addition, the higher intensity toroidal currents will probably be mutually attracted by stronger forces.

# *3. 1. Successive processes in a typical 22-year solar cycle*

We shall exhibit a reasonable evolution cycle of processes incorporating the butterfly 11-year sunspot cycle and the reverse mechanism of the magnetic field in a self-consistent way. Given that all the contributing processes (based simultaneously on charges, currents and magnetic fields) have been already studied in detail in the preceded section, we continue producing an overall view using six schematic diagrams emphasizing the main aspects of the model unifying different parts of the solar puzzle. Our synthesis will have, in this section, inevitably a synoptic character; after the end of this "skeleton description", we shall further clarify and discuss several worth noticing issues. Well-known observations concerning the solar differential rotation and the emergence of sunspots in pairs, obeying a well determined law with respect to their polarities, are all taken into account. Thus, the sketch of **Fig. 9** shows only six fundamental snapshots taken from an entire 22-years solar cycle; the three left (right) column panels correspond to the 11–years cycle with northward (southward) dipole magnetic field. Briefly, we shall go throughout each panel pointing out its basic functionality.

### *Figure 9a, a view above the equatorial plane*

The two homocentric circles denote the surfaces of the RZ and photosphere projected over the equatorial plane. The initial "seed field" $\mathbf{B}_o$ (arbitrarily chosen being northward directed here) initiates a process of charge separation in the Convection Zone; electrons move inwards forced by the Lorentz force, $\mathbf{F}=-e\mathbf{u}_e\mathbf{x}\mathbf{B}_o$, where $\mathbf{u}_e$ is the electron velocity resulting from the (eastward) rotational speed of the Sun. The RZ (assumed as a perfect conductor) is under negative charging $(Q^-)$, whereas a layer beneath the photosphere is developing a positive charge $(Q^+)$; the overall charge $(Q^++Q^-)$ is zero. The inward electron motion is particularly increased at low latitudes; as the latter is dictated from the solar differential rotation. The surface charge of the RZ (sub-photosphere layer) creates an azimuthal westward (eastward) directed current, $J_\varphi$. The total magnetic field in the Convection Zone is a superposition of both azimuthal currents, $B_z=B_{RZ}+B_{photo}$. The $J_\varphi$ of the RZ generates the northward directed dipole field, $B_{dip}$, outside the Sun (look at the subsection 2.13). That is, outside the Sun $B_z=B_o+B_{dip}\approx B_{dip}$. However, given that the $B_{dip}$ survives for a long time (i.e., about 11-years) and the $Q^-$ could not monotonically increase, a discharge process of the RZ-sphere must be at work, too. Electrons must escape radially outwards from the poles of the RZ.

### *Figure 9b, a meridional plane view*

The electrons flow radially outwards from the equipotential surface of the RZ through the polar areas characterized by an electrostatic field almost in parallel to the magnetic field. The resistivity of the Convection Zone roughly increases in proportionality to its radius. The prevailing repulsive electrostatic force, over the polar regions, further spreads out the negative charge escaping from the RZ. The electrons move towards the photosphere layer and may diffuse equatorwards; in this way, two meridional current cells are probably developed, one in each hemisphere. Obviously, the situation may not be similar to a typical current loop.

### *Figure 9c, a meridional plane view*

The radially outward escaping electrons, within the Convection Zone, set up an inward current flow, $J_z$. This current generates the toroidal "seed magnetic field", $B_{tor}$, for each Torus; westward (eastward) magnetic field in the northern (southern) hemisphere. Each negatively charged Torus, while drifting equatorwards constitutes the "fast dynamo engine" of the Sun. The very intense toroidal currents, $J_{tor}$, are always westward directed and the associated strong poloidal magnetic fields attract more and more electrons, gaining their entry into the Torus entity. This positive feedback, beside the zero resistivity within the Torus, is really creating an ideal "fast dynamo" action. Each Torus steadily shifts equatorward and finally disintegrates at the equatorial plane. The combined action from the two toroidal currents plays the key role in reversing the dipole magnetic field. ***There is an instant at which the magnetic field, being tangential to the RZ, reverses its sign.*** The switching from northward to southward directed magnetic field occurs when the two tori approach each other and the produced southward field in low latitudes halts the electron supply to the RZ. Then the southward field constantly grows and the low latitude electrons are removed farther outwards from the RZ; finally, the RZ is switched to positive charging ($Q^+$, sketch 9d).

### *Figure 9d, an equatorial view*

Once inaugurated, the positive charging of the RZ will continue for about 11-years. The $Q^+$ charge will essentially support the southward directed dipole magnetic field (outside the photosphere). The (related with the $Q^+$ charge) eastward current, $J_\phi$, flowing over the surface of the RZ forces (via its poloidal magnetic field) the Convection Zone electrons to move outwards; in this way, a negatively charged layer beneath the photosphere is formed. Actually, we have two azimuthal anti-parallel currents producing the overall southward directed magnetic field within the Convection Zone; outside the polar regions of the Sun, the prevailing field is that corresponding to the $J_\varphi$ of the RZ. Most importantly, the sub-photosphere current layer becomes exceptionally thin and the charge density extremely high; thus, the repulsive electrostatic forces may move the electrons poleward (although remaining inside the thin sub-surface layer).

### *Figure 9e, a view of meridional plane*

The southward directed magnetic field, within the Convection Zone, steadily charges positively the RZ and negatively the sub-photosphere layer. The latter is more emphatic at low latitudes where higher azimuthal velocities are developed (due to the solar differential rotation). Over the polar regions, electrons move radially inwards, along the North-South Z–axis, forced by attractive electrostatic forces. In these regions, there is none magnetic force preventing this motion. Finally, the electron circulation apparently sets up a meridional current flow; the situation looks like comprising two meridional current cells having opposite flow directions in comparison with those shown in the preceded 11-year cycle (sketch 9b).

### *Figure 9f, a meridional plane view*

The radially outward flowing current, $J_z$, generates a toroidal magnetic field evolving to a mature Torus, in each hemisphere. However, in this phase the toroidal magnetic field is eastward directed in the northern Torus and westward in the southern one. And this outcome may remind us the concept based on the MHD approach that the magnetic field lines (which are southward directed here) are coiled around the Sun because of its differential rotation. In this phase, the poloidal magnetic field of each Torus repeals an ever increasing number of electrons outside the Torus. In this fashion, the toroidal currents are gradually intensified, while the positively charged tori progressively move equatorwards and finally disintegrate over the equatorial plane. The northward directed toroidal magnetic field prevails in the region close to the tachocline and finally the sign of charge reverses on the surface of the RZ from $Q^+$ to $Q^-$. The drift of tori essentially produces the sunspot butterfly diagram lasting for about 11–years and concerning the sunspot activation centres.

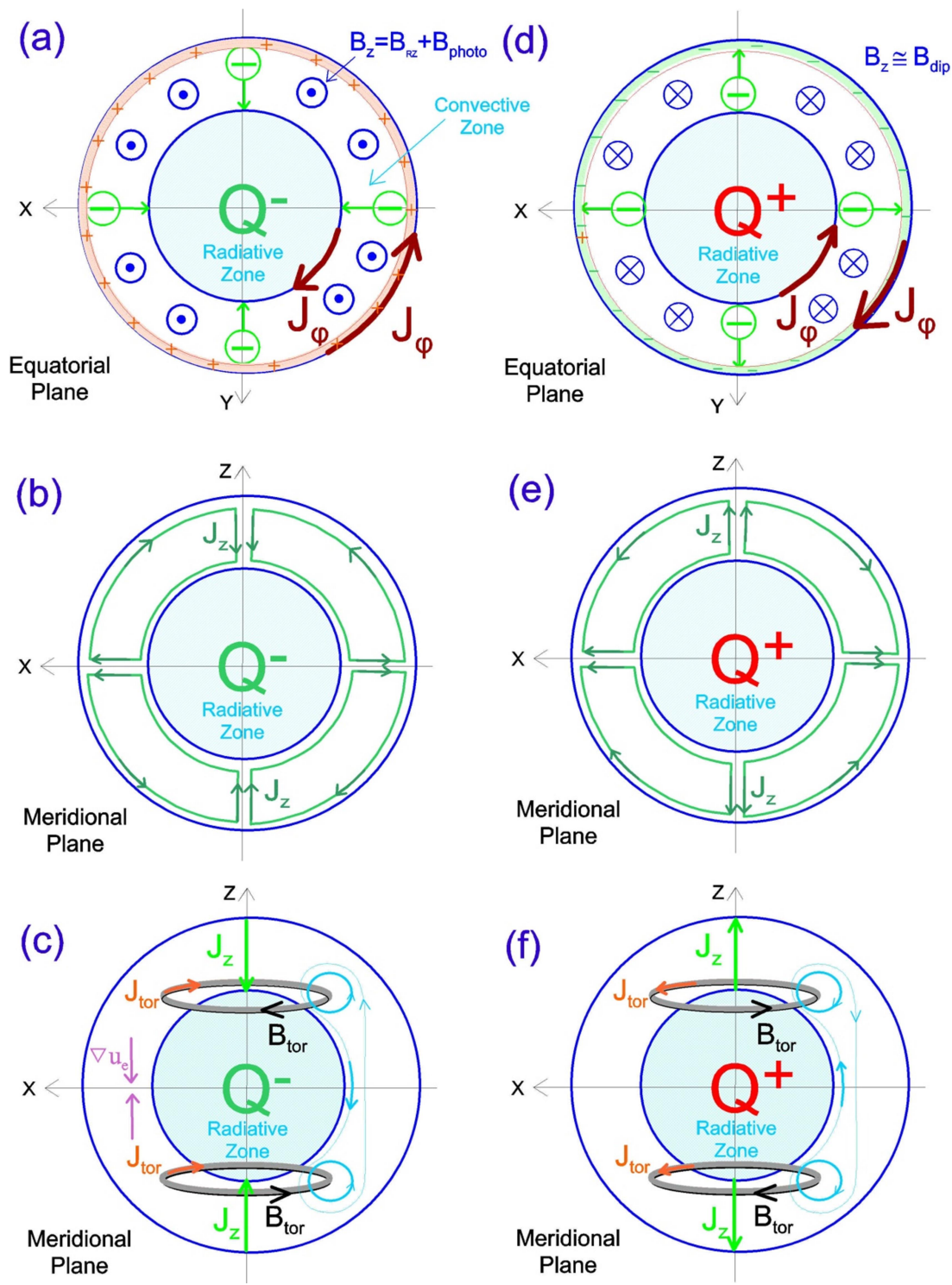

**Fig. 9.** The azimuthal currents projected over the equatorial plane for the case of northward (a) and southward (d) directed dipole magnetic field. From (a) to (d) the time period is about 11 years, when the dipole magnetic field is switched from northward to southward and the RZ flips from $Q^-$ to $Q^+$. From (d) to (a) the time is also about 11 years (i.e., the next sunspot cycle), when the dipole magnetic field is switched from southward to northward and the RZ flips from $Q^+$ to $Q^-$. The radially inward (c) and outward (f) flowing $J_z$-current generates the toroidal magnetic field resulting to a mature Torus in each hemisphere. In successive cycles the toroidal currents and magnetic fields are oppositely directed.

## *3.2. Additional clarifications*

**Figure 9a (9d)** shows the azimuthal currents projected over the equatorial plane, for the case of northward (southward) directed dipole magnetic field. From (a) to (d) the time period is about 11 years (i.e., one sunspot cycle); then the dipole magnetic field is switched from northward to southward while the RZ flips from $Q^-$ to $Q^+$. From (d) to (a) the time period is also about 11 years (i.e., the next sunspot cycle), when the dipole magnetic field is switched from southward to northward and the RZ flips from $Q^+$ to $Q^-$. The period of the positive charging $Q^+$ ($Q^-$) of the RZ lasts about 11 years and is associated with southward (northward) magnetic field. The $Q^-$ ($Q^+$) of the RZ is essentially related to a negatively (positively) charged Torus.

The meridional view of the northward (southward) magnetic field associated with $Q^-$ ($Q^+$) is separately sketched in **Fig. 10a (10b).** The magnetic field in the Convection Zone is shown here as superposition of the two azimuthal currents. For instance, in **Fig. 10a** the electron current flows westward over the RZ and the ion current flows eastward in a sub-photosphere layer. Exactly the opposite electron circulation is depicted in **Fig. 10b**; the sub-photosphere layer (RZ) is related to a westward (eastward) electron (ion) current. In Fig. 10a, the low latitude electrons are inward attracted through the prevailing magnetic forces in tachocline. In contrast, the polar region electrons are pushed away (from the RZ) by the dominant electrostatic forces. We have to pay attention that, in **Fig. 10b**, the electrons are accumulated in a very thin layer beneath the photosphere, and a potential difference is developed across the low latitude Convection Zone. These electrons either remain there or move polewards by repulsive electrostatic forces if the partially ionized local plasma loses its ability to screen out the penetrated negative charge. In any case, electrons will certainly flow radially inwards in the polar region.

In the sketch of **Fig. 9c**, the two toroidal structures approach each other, while the dipole field is northward directed. The tori are immersed within the tachocline (transition layer). The toroidal currents flow westwards and produce poloidal magnetic field, which is southward directed adjacent to the RZ. As the time proceeds, there is a moment when the adjacent to the RZ prevailing magnetic field is switched to southward, at low latitudes. Then, the negative charging of the RZ stops, and as the electrons begin to escape radially outwards, at low latitudes, the RZ will inevitably change its sign of charging. The positive charging will continue throughout the gradual disintegration process of both tori and the systematic accumulation of electrons in the sub-photosphere layer.

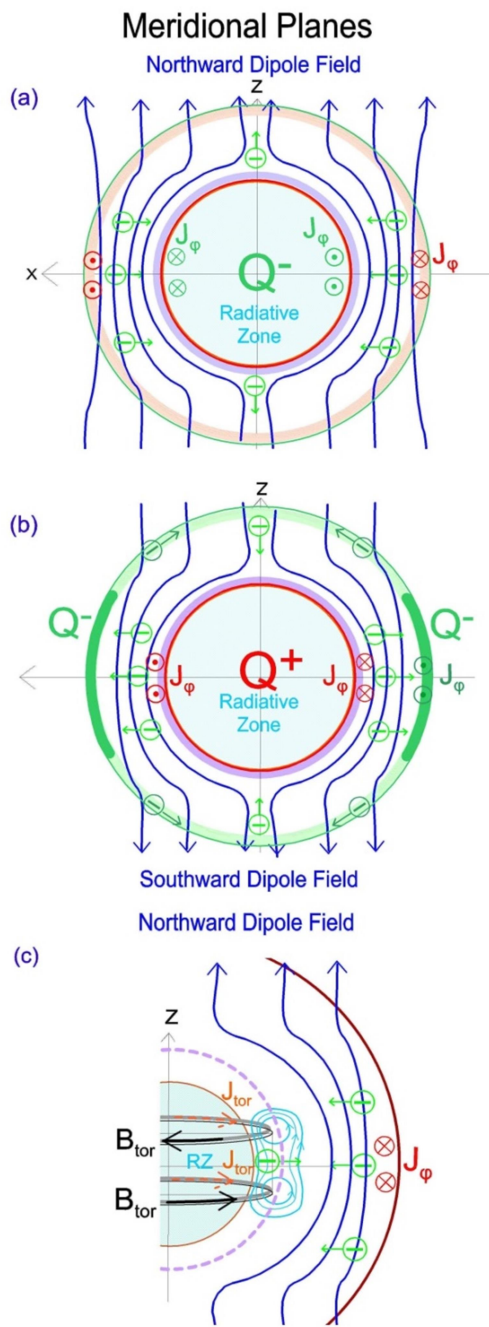

**Fig. 10.** Meridional view with northward (a) and southward (b) directed dipole magnetic field associated with Q⁻ and Q⁺ charges on the surface of the RZ, respectively. The Convection Zone magnetic field is shown here as superposition of two azimuthal currents. Either the electron current flows westwards over the RZ, and the ion current flows eastwards in a sub-photosphere layer (a) or the ion current flows eastwards over the RZ and the electron current flows westwards in a sub-photosphere layer (b). (c) As the two (negatively charged) toroidal structures approach each other, there is a moment when the prevailing magnetic field adjacent to the RZ is switched to southward and the RZ changes its sign of charging from negative to positive.

# *4. Discussion*

## *4.1. The solar 3D circuit*

The sketch of **Fig. 11** is a preliminary approach to a "3D solar circuit" being symmetric with respect to the equatorial plane. One can discern three distinct entities (i.e., three mutually coexisting sub-circuits) related to (a) the Convection Zone (**CZ**), (b) the Radiative Zone (**RZ**) and (c) the **Torus**. The latter is the only one complex originated, matured and vanished every ~11 years. Moreover, the capacity $C_{CZ}$, the resistance $R_{photo}$ and the inductances $L_{RZ}$ and $L_{CZ}$ are related to the circulation path of the meridional current within the CZ. The CZ (associated with the upper part of the whole circuit) is related to

1. a charge separation process producing a huge almost "cylindrical capacitor" extended all the way from the tachocline up to the photosphere and establishing the meridional current flow (within the CZ),

2. an electron circulation within the CZ (due to the voltage difference developed across the capacitor) changing its flow direction every ~11 years,

3. a **slow dynamo action,** since both the CZ and the involved sub-photospheric layer have increased resistivities,

4. an **ω-effect action,** since the large-scale poloidal magnetic field and the meridional current in the CZ produce the toroidal field of each Torus,

5. the two azimuthal currents, flowing the one along the Torus and the other over the RZ-surface. Both of them are generated via the rotational-mechanical energy of the Sun. In the sketch, the solar rotation is shown using two parallel-thick lines signifying the coupling between the CZ and RZ, and CZ and Torus, and

6. a circuit having its own capacitor $C_{CZ}$ and inductors ($L_{RZ}$ and $L_{CZ}$). Therefore, this circuit has its own **resonance frequency corresponding to T≈22 years**. Apparently, via the $L_{RZ}$ and $L_{CZ}$ the meridional current is coupled with the two azimuthal currents related to the RZ and the two tori.

The RZ is associated with

1. an azimuthal current (due to its positive or negative net charge) flowing over the surface of the RZ, and

2. **the weak dipole field** reversing its polarity every ~11 years.

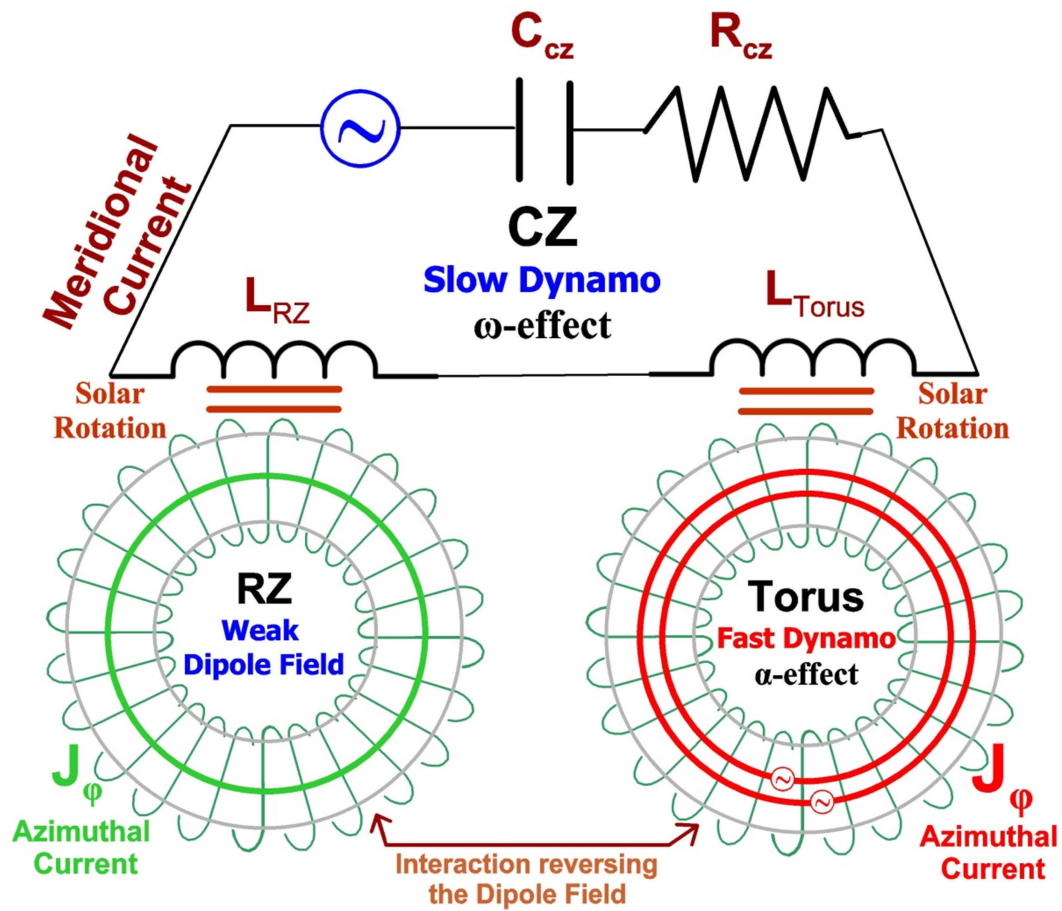

**Fig. 11.** A "solar circuit" (corresponding either to the northern or southern hemisphere) composed of three mutually coupled sub-circuits which are related to (a) one meridional current flowing in the Convection Zone and (b) the two azimuthal currents flowing along the surface of the Radiative Zone and the Torus. Symbols RZ, CZ, Torus, $C_{CZ}$, $L_{CZ}$ and $L_{RZ}$ denote the Radiative Zone, the Convection Zone, the toroidal complex, the capacitance and the two inductances of the meridional circuit. The inductances mutually couple the two azimuthal currents of the Torus and RZ. The Torus (characterized by an inherent zero resistivity) generates the "fast dynamo action" and the "α-effect action" via the toroidal current and its own poloidal magnetic field. The azimuthal current of the RZ essentially produces the weak dipole magnetic field that interacts with the poloidal field of the Torus and reverses the magnetic field polarity. The sign of charge of the Torus (and the RZ alike) determines the electric current flow direction.

The "Torus structure" is associated with

1. the **fast dynamo action** given that the azimuthal (toroidal) current flows within the Torus with zero resistivity; the current flow direction "swings back and forth" in each 22-year solar cycle,

2. the **a-effect action,** due to the enormous toroidal current intensities, producing exceptionally strong poloidal magnetic fields,

3. an extraordinarily stable structure with none destructive force, except for the time period near to the end of the 11-year cycle, when apparent symptoms of disintegration appear,

4. the strong poloidal magnetic fields interacting with the weak pre-existing poloidal fields of the RZ and finally reversing the polarity of the magnetic field, and

5. the strong poloidal magnetic fields acting in a sub-photosphere layer (characterized by an intense ω-shear) and essentially producing the sunspot groups and the CME events.

## *4.2. Torus-Sunspot group relationship*

The differential rotation of the Sun (inferred from global helioseismology) is of paramount importance in the sunspot formation process. The Torus dynamics in parallel to the differential rotation essentially generates the sunspot pairs. On the issue of differential rotation, two features have been already cited in subsection 2.2; here, we would like to mention one more outstanding feature playing decisive role in the sunspot formation. The latter is described as follows: There is another transition layer (in addition to the tachocline) near the top of the Convection Zone with comparatively large radial shear in the angular velocity ω. At low and mid-latitudes there is an increase in the rotation rate immediately below the photosphere, which persists down to r ≈ 0.95 $R_{sun}$; that is, this surface layer has thickness ~5.5 $R_E$ (earth radii). The angular velocity variation across this layer is roughly 3% of the mean rotation rate and according to the helioseismic analysis of Corbard and Thompson [19] ω decreases within this layer approximately as $r^{-1}$. At higher latitudes the situation is less clear.

The two toroidal entities, while performing their equatorward shift, directly affect the top layer of the Convection Zone, which is probably responsible for the formation of sunspot pairs with different polarities and sunspot groups alike. ***The Torus strong poloidal magnetic field, resulting from its toroidal intense current, essentially dictates the butterfly diagram in an 11-year cycle.*** As we shall analytically exhibit bellow ***an Emerging Flux Region (EFR) is initially formed in a layer located 5.5 $R_E$ beneath the photosphere***. From that birth place it grows from inside to outside, penetrating the photosphere and emerging as an Ω-shaped loop with separate polarities. Initially, the EFRs may be oriented almost at random, but they rotate and after about one day most of them achieve an orientation nearly parallel to the equator.

The Torus poloidal magnetic field prevails against any pre-existing magnetic field in the Convection Zone (CZ). Thus, the region close to the photosphere is charged positively (negatively), because the electrons of the CZ move radially inwards (outwards) affected by the Lorentz-magnetic force related to a northward (southward) directed poloidal magnetic field. When the Torus current flows westwards, the number density of positive charge dramatically increases at those latitudes mainly influenced by the Torus poloidal magnetic field. Especially, near the top of the Convection Zone (i.e., at $r=0.95$ $R_{sun}$) much higher amounts of positive charge are accumulated due to the large radial $\omega$-shear. And these locally accumulated sub-surface charges set up an eastward (azimuthal) current flow, $J_{\phi}$. Additionally, since the rotation rate decreases monotonically toward the poles by about 30%, the charge density is always characterized by a gradient pointing equatorward.

The schematic of **Fig. 12**, showing the RZ-surface and the photosphere as two vertically directed planes, illustrates the whole scenario for the case in which the Torus is negatively charged. The prevailing magnetic field beneath the photosphere is northward directed, implying that the positive charge density is threaded by northward magnetic field lines. Thus, if a "parcel" of positive charge is azimuthally restricted, then the resulting local structure would (at least morphologically) be a magnetic flux rope (MFR) like structure. And the latter is a very realistic hypothesis since the Torus azimuthal current might be highly variable even at neighbouring longitudes. From its very generation mechanism this current does not obey any "current continuity" equation; it results from the inhomogeneously distributed net negative charge moving with the solar rotational speed. In this way, a local enhancement of the Torus azimuthal charge will immediately produce an increased positive charge in the sub-photosphere layer. Eventually, a charged parcel with an almost cylindrical magnetic structure is formed, while a potential difference is applied on the two ends. Most importantly, an outward pointing magnetic force could raise this entity outside the photosphere forming a pair of sunspots with opposite polarities. And the whole emerging structure will certainly continue carrying an amount of net positive charge. The magnetic force results from the vector sum of the magnetic field due to the azimuthal–eastward current (of the positive charge) plus the northward directed magnetic field. In turn, the lower latitude sunspot, corresponding to higher local charge densities, might produce FACs within the emerging flux tube. Especially outside the photosphere (characterized by much lower plasma densities) the FACs are probably allowed to flow, twisting the initially straight magnetic field lines. Thus, one can reasonably argue that the emerging flux tube will progressively become much more stable since the growth of the flux tube seems to be associated with an ever increasing poloidal magnetic field (due to the progressively intensified FAC of tube). The whole scenario will change appropriately with southward directed magnetic field in the Convection Zone.

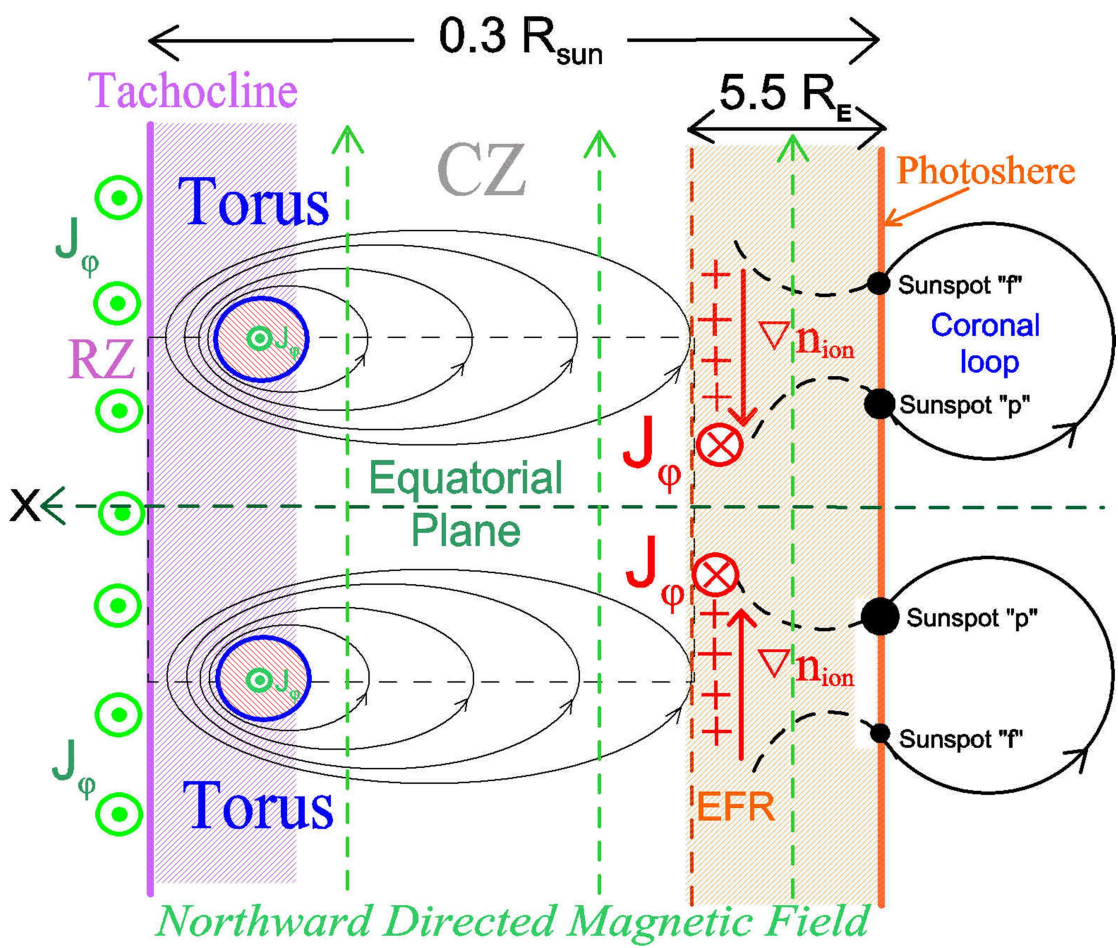

**Fig. 12.** In this schematic the RZ-surface and the photosphere are shown as two vertically directed planes. The Torus currents are sketched flowing westwards. At low and mid-latitudes there is an increase in the rotation rate immediately below the photosphere which persists down to $r \approx 0.95\ R_{sun}$. An Emerging Flux Region is initially formed in a layer located 5.5 $R_E$ beneath the photosphere. From that birth place, it grows from inside outside, penetrating the photosphere and emerging as an $\Omega$-shaped loop with separate polarities.

## 4.3. Morphological asymmetry between the leading and following sunspots

Sunspot pairs (or bipolar groups) are tilted, with the leading sunspots closer to the equator than the following. The tilting phenomenon of sunspot pair is known as Joy's law, while the tilting angle increases with latitude (e.g., [20]). Since sunspots generally form in magnetically-linked bipolar groups; the magnetic axis of the sunspot group is usually slightly inclined to the solar east-west line, tilting from $3^{\circ}$ near the equator to $11^{\circ}$ at latitude $\pm 30^{\circ}$, with the preceding polarity spot (i.e., the "p" one) being slightly closer to the equator. In a bipolar group of sunspots, if the axis is highly tilted initially, the group will tend to rotate until the axis achieves a direction more parallel to the equator.

An intriguing property of solar active regions is the asymmetry in morphology between the leading and following polarities (i.e., the "p" and "f", respectively). The leading polarity of an active region tends to be in the form of large sunspots, whereas the following polarity tends to appear more dispersed and fragmented; moreover, the leading spots tend to be longer lived than the following (e.g., [21] and the review of Fan, [11]).

In the context of this work and according to the preceded subsection 4.2, each sunspot pair (produced by a negatively charged Torus) is associated with a regional positive charge distributed in a well-determined sub-photosphere layer. Moreover, since an azimuthal speed gradient is always directed equatorward, a greater charge density is anticipated at the loop foot of the leading sunspot. A potential difference is developed in each sunspot pair. Consequently, it follows that, inside an emerging tube, the FACs are stronger close to the leading sunspot. The poloidal magnetic field becomes preferentially stronger close to the leading sunspot and ***it finally gives rise to an effective asymmetry with stronger total magnetic field in the leading leg***. The FACs flowing along the loop twist the magnetic field and the highly twisted magnetic flux tubes may be associated with flare productive active regions. The whole scenario for an emerging coronal loop is illustrated in **Fig. 13**. And under this perspective one can easily comprehend why the equatorward located sunspot is usually larger, better developed and with relatively longer lifetime. The collapse of this coronal loop-structure will probably produce a solar flare, as we propose in the next subsection. About solar flares one can read the reviews, for instance, of Shibata and Magara [22] and Benz [23].

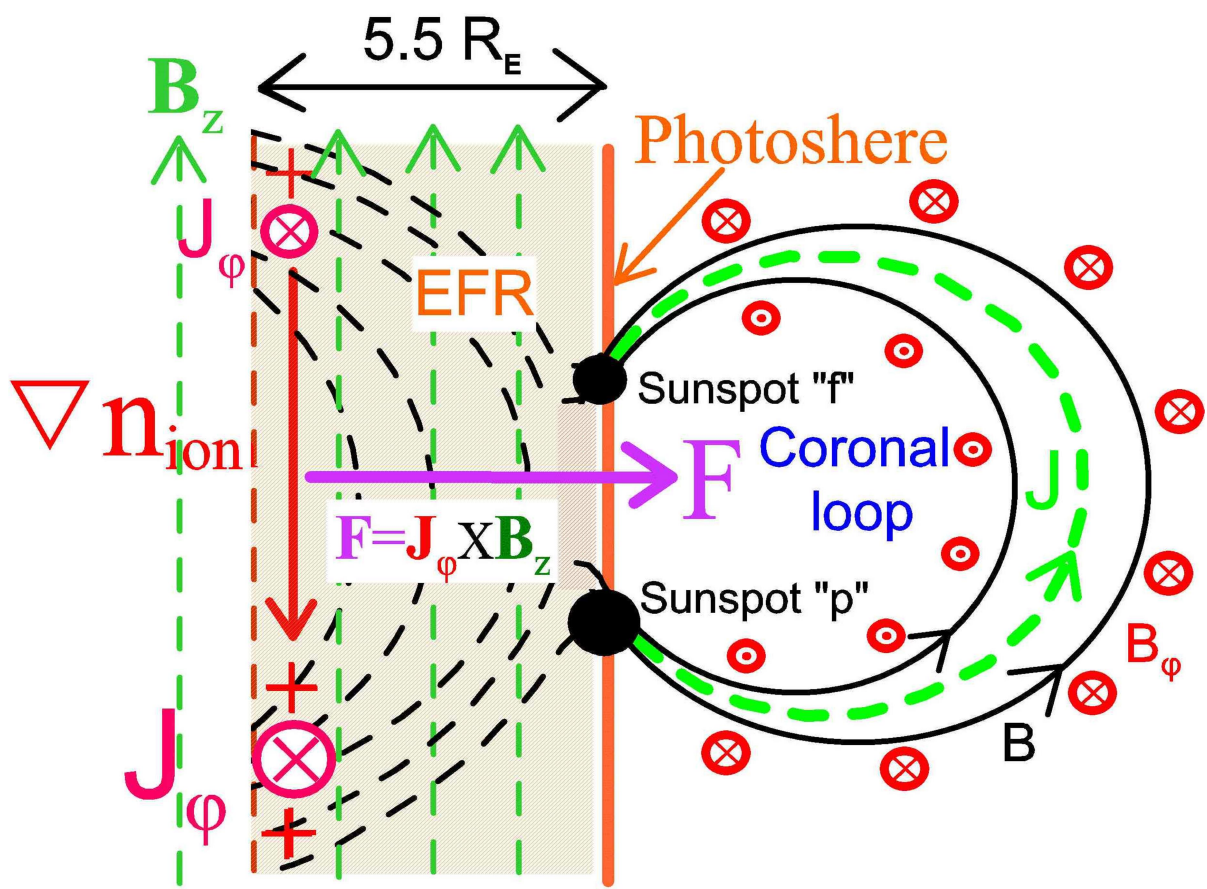

**Fig. 13.** For a negatively charged Torus, the subsurface positive net charge is mainly developed at r=0.95 $R_{sun}$; inside this plane there is no surplus positive charge. The net charge sets up the azimuthal current $J_{\varphi}$ which along with the locally northward directed magnetic field $B_z$ (black-dashed lines) raises the magnetic tube outwards. The emerging magnetic tube probably develops FACs outside the photosphere, since a potential difference is applied on it. The resulting poloidal magnetic field, $B_{\varphi}$, further increases the stability of the already formed coronal magnetic loop. The sunspot located equatorward is usually larger, better developed and with relatively longer lifetime. For a positively charged Torus, the directions of $J_{\varphi}$ and $B_z$ are reversed; the magnetic tube raises outwards, too.

## *4.4. Coronal loop collapse*

We hypothesize (in the case of a negatively charge Torus) that a stable coronal loop structure is essentially composed of a poloidal magnetic field tube which is gradually evolving in a loop shape initially on a plane perpendicular to the Sun's East-West direction. Such a loop, placed at the western edge, is shown in **Fig. 14a**. In addition, if the whole loop is subject to a cyclonic motion (like that occurring in the Earth's atmosphere) due to the solar differential speed, then the loop would rotate and be adjusted almost in parallel to the East-West direction (represented by the X-axis in **Fig. 14b**). However, the new orientation probably triggers a destabilizing mechanism leading to a massive collapse of the loop. The initially azimuthal current (i.e., the $J_\varphi$), generated by a layer of the emerging magnetic loop in the sub-photosphere which is positively charged, will become an almost field-aligned current after the rotation. As a result, the previously upward directed force, supporting the huge arch, vanishes. Consequently, a solar flare can be expected to occurs, accelerating huge amounts of electrons. This phenomenon will be the subject-matter of the next subsection.

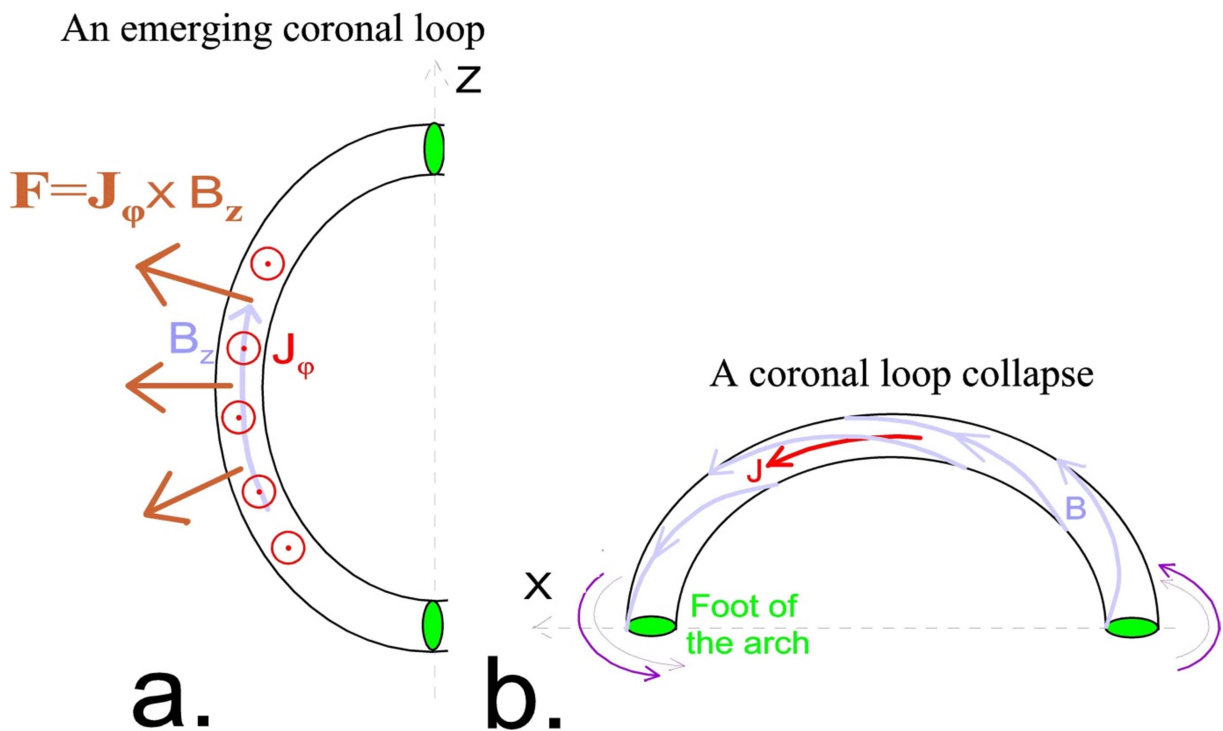

**Fig. 14.** (a) An emerging coronal magnetic loop (on the western edge of the Sun with a negatively charged Torus) is initially orientated in parallel to a meridional plane. (b) A cyclonic motion turns the whole loop almost in parallel to the equatorial plane; consequently the initially azimuthal current, $J_\varphi$, becomes field-aligned and the force $F=J_\varphi \times B_z$ raising and supporting the loop vanishes. The latter constitutes a disastrous process leading to a solar flare associated with massive electron acceleration.

## 4.5. The way in which $10^{36}$ electrons may be accelerated by a solar flare

The most difficult problem in explaining the behaviour of impulsive flares that theorists face is the charged particles acceleration to high energy in so short time and so limited space. Nearly the entire population of electrons in a flaring loop (i.e., $\sim 10^{37}$ particles) accelerates to energy as high as 100 keV within one or two minutes (e.g., [24]). The bulk of fast electrons spiral down the field lines in the legs of the loop and emit polarized (gyrosynchrotron) microwaves along the way. A large fraction from the flare energy is released as the kinetic energy of fast electrons (e.g., [25]).

In this subsection, we shall try to address the problem of the energetic electrons striking into the chromospheric "footpoint sources"; while a scenario for "a top-loop process" accelerating electrons is suggested in the next subsection. We shall propose an acceleration mechanism potentially taking place during the collapse of a coronal loop and especially within the two plasma columns. Persistent and very intense inductive electric fields are developed along the magnetic field lines from the falling plasma. ***Each plasma column seems to act like a powerful electromagnetic tornado.***

In a first place, we assume that a vertical, almost cylindrical structure of plasma, begins to collapse downwards (subsection 4.4) affected by the gravitational force. In a second place, we assume that the whole coronal-loop structure inherently contains an amount of surplus positive charge (in the case of a negatively charged Torus) and is characterized by a helical magnetic field topology (i.e., a field having a $B_z$-longitudinal component along the axis of the tube, plus a $B_\varphi$-azimuthal one). The downward moving ions with velocity V (**Fig. 15**), during the downfall of the plasma column, establish (a) an intense downward directed current, and (b) a radially inward motion forced by the Lorentz force $qVB_\varphi$. Additionally, the ion acceleration due to gravity gradually increase the vertical current density, $J_z$, and the tube reduces its cross-sectional area becoming narrower close to its bottom (i.e., the tube is conically shaped). Then, the azimuthal magnetic field further becomes greater, leading to a higher vertical current and so on. Therefore, an explosive process is initiated that dramatically increases the vertical magnetic field component, $B_z$. The latter (i.e., the abrupt change of the $dB_z/dt$ rate) produces extremely intense and magnetic field-aligned azimuthal electric fields, $E_\varphi$. Consequently, a large fraction of the electrons from the disintegrating cylindrical structure would be easily accelerated. To conclude, the inductive electric fields are essentially due to the ions of the surplus positive charge, existing within the plasma column and leading to massive electron acceleration.

Certainly, according to the "standard flare model" (e.g., look at the review of Mann, [25]) electrons are first accelerated at the magnetic reconnection site, and then they travel down and emit hard X-rays on the chromosphere; that is, the primary energy release and the electron acceleration take place in the corona. In contrast, in agreement with the view adopted here, although the reconnection is probably taking place, however, the main acceleration mechanism is realized otherwise. The reconnection is mainly a process of magnetic field reconfiguration.

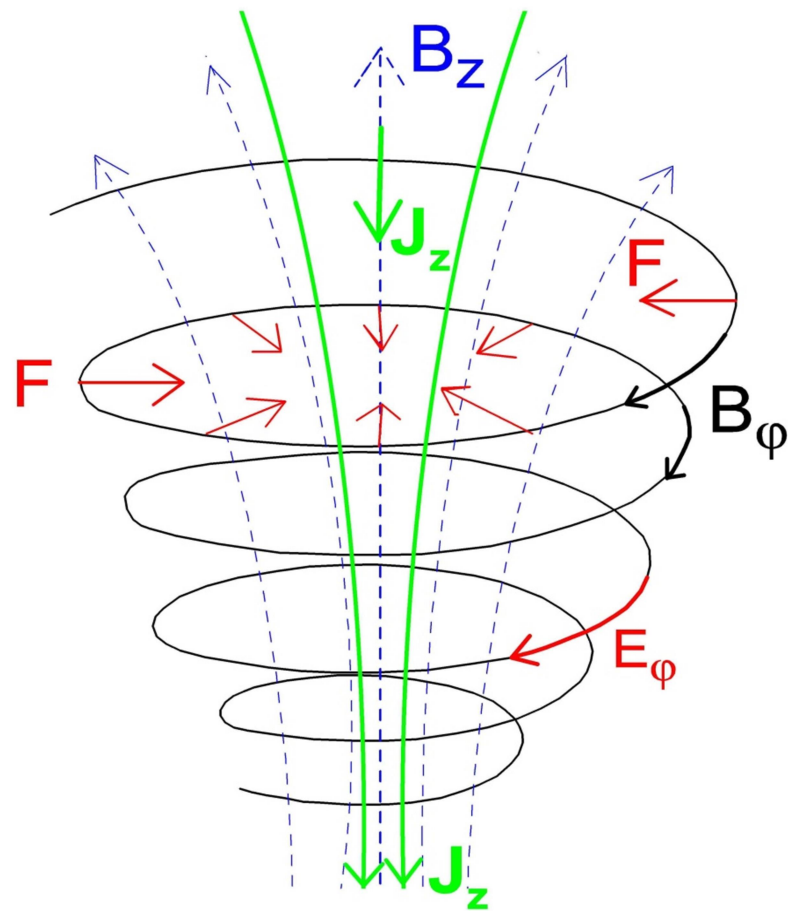

**Fig. 15.** The plasma column with cylindrical geometry, having as main ingredient a surplus positive charge, collapses under the gravitational force $F_g$. The downward falling ions with velocity V are forced to move inward by the magnetic force $F=qVB_\varphi$. The longitudinal current density $J_z$ and the magnetic field $B_z$ increase explosively producing extremely high inductive electric fields ($E_\varphi$) along the azimuthal component ($B_\varphi$) of the magnetic field. A similar scenario corresponds to the collapse of a negatively charged plasma column (in the case of a positively charged Torus).

## *4.6. Origination of CMEs*

The coronal mass ejection (CME) event (e.g., [26]) is the most powerful one occurring outside the photosphere, while remaining tightly associated with processes taking place in the Sun's interior and more specifically with those caused by the two toroidal structures.

The CME event is a large scale phenomenon. In contrast, the commonly observed solar flares are related to small scale coronal magnetic loops (as those studied in the four preceded subsections). In the CME situation, we assume again that in the sub-photosphere layer (developed ~5.5 $R_E$ beneath the photosphere) a positive charge is accumulated (in the case of a negatively charged Torus). However, in this case the ions have much higher ion densities and are distributed along an extremely lengthened area. Such an area will evolve into a solar filament (prominence) extended up to about a solar radius. This prominence would be produced by a sharp (spatial) spike (i.e., a sharp local rise followed by a sharp decline) in the Torus current. All the other coronal loops may be dictated by small fluctuations of the Torus current and would be related to small scale charge gradients in the sub-photosphere layer. In contrast, the CME is directly generated from an extreme and locally restricted intensity of the Torus current.

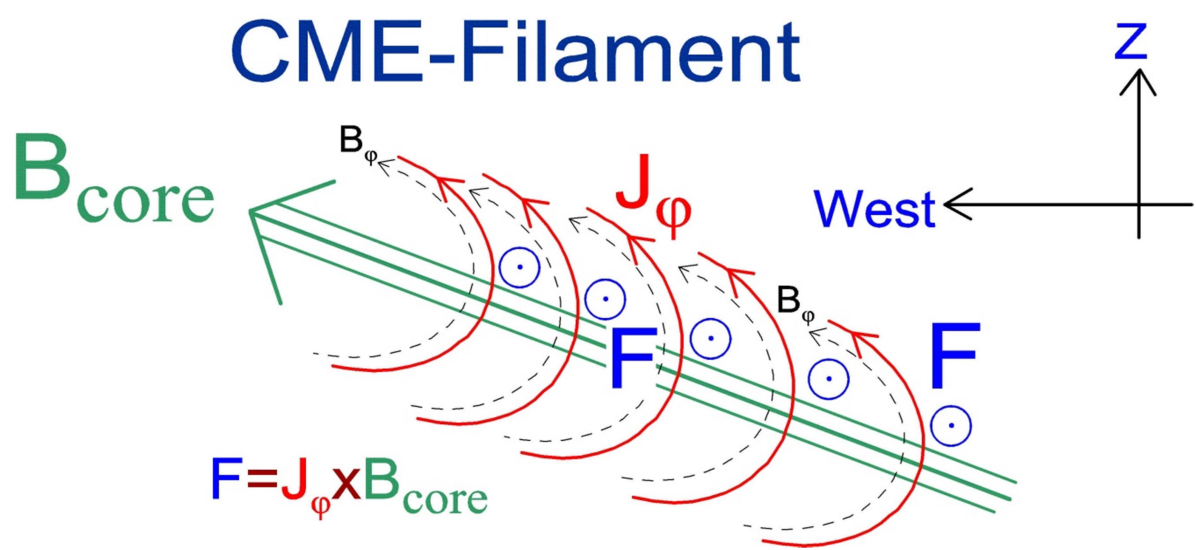

**Fig. 16.** CME, being a gigantic MFR, violently expands upwards by the magnetic force $J_\varphi XB_{core}$. The CME is probably composed of "small scale MFRs" having their own cores along which flow the $J_\varphi$-FAC producing the central large scale "core magnetic field", $B_{core}$.

Therefore, in this perspective, a CME is essentially composed of many magnetic loop-like tubes (like those studied in subsection 4.4) placed the one beside the other, while being northward directed along the Z-axis (**Fig. 16**). In this way, the $J_\varphi$-FACs flowing along each loop will probably build up a strong "core magnetic field", $B_{core}$. And the core will probably form an "inclination angle" with the equatorial plane, as the latter is frequently observed in filaments. Undoubtedly, an initial north-south charge distribution would be seriously modified by the differential solar rotation. Finally, the upper portion of the filament would violently expand outwards by the magnetic force $\mathbf{F} = \mathbf{J_\phi} \times \mathbf{B_{core}}$. And the upward rising of the huge structure potentially leads to super-Alfvénic jets, which in turn produce fast magnetosonic shocks producing a large amount of accelerated electrons (i.e., energies greater than 30 keV).

## *4.7. A comment on Stellar Dynamos*

For main-sequence G and K stars it was found that ''the periods of the activity cycles increase proportional to the rotational periods, along "two distinctly different sequences", the "active" and "inactive sequence" with cooler and more slowly rotating stars" ([27, 28]). According to the mainstream understanding of astrophysical magnetic fields, their generation mechanism is usually based upon a self-excited dynamo, which involves flows that can amplify a weak "seed magnetic field" exponentially fast. It is usually assumed that the field is created by the interaction of rotation, convection and the angular velocity gradients in the Convection Zone.

In the context of this work and its dynamo model, the "star activity period" is mainly depended on (a) the two toroidal currents formed inside the star and placed symmetrically with respect to the equatorial plane, and (b) the surface azimuthal current flowing over the RZ. Consequently, the above mentioned proportionality may easily result since all the currents are directly related to the angular velocity, $\omega_{star}$; the same is valid for the growth time of the net charges accumulated in the RZ and the two tori alike. With an enhanced $\omega_{star}$, the Torus current grows faster and the magnetic force, attracting (or repelling) electrons, grows even faster. Furthermore, in our perspective, the azimuthal currents are obviously depended on an additional parameter; the star radius, $R_{star}$. For instance, the azimuthal current flowing over the RZ is $J_\phi = \rho \omega_{star} R_{star} \sin\theta$, where $\theta$ is the latitude angle. As the $R_{star}$ increases, all the currents inevitably increase and the same does the Torus current (mainly affecting the star's activity period). Therefore, we make a suggestion that beyond the angular velocity there is an additional parameter of crucial importance, the $R_{star}$. And, one may hypothesize that the above mentioned "two distinct different sequences of stars" are related to the two independent parameters. Certainly, the validity for this claim must be checked and observationally supported by future research efforts.

Moreover, it was found that along each sequence, the number of rotation periods per activity cycle is nearly the same, but the numbers are different for the different sequences, indicating that probably different kinds of dynamos are working for the stars on the different sequences (e.g., ([28]). In the context of this work, two kinds of dynamos are not unavoidably needed; conversely, the one mechanism may split into two (or more categories) as two are the chief parameters. Either the $\omega_{star}$ or the $R_{star}$ plays the dominant role.

## 4.8. Magnetars may be generated from the exotic "solar Torus" properties

In the presented dynamo model of the Sun, the convective motions, which are not anticipated in a neutron star, do not play any significant role producing the fast dynamo action. In addition, the high electron densities presumably developed (for instance) in a negatively charged Torus may be fitting with the pressures characterizing the outer layer of a neutron star. Therefore, the fundamental idea of this work may be applicable even in the case of a magnetar. After the process giving birth to a neutron star, it may carry a negative charge in a thin layer around the spheroid determined by the relation $u_e = c'$. We could hypothesize that a large fraction of the electrons from the outer crust of the neutron star, having thickness 0.3-0.5 km, would be accumulated in this thin layer. The electron rotational speeds would be only a few times higher than those reached in the Sun's situation; for instance $u_e \approx 14$ kms$^{-1}$, if the rotation period is 5 s (for a magnetar) and the radius 11 km. Nevertheless, in a neutron star there are extreme density values of $\sim 4 \times 10^{14}$ g/cm$^{-3}$. That is, a density higher $\sim 10^{15}$ times the density value of the Sun; given that, for instance, the tachocline has density 0.2 g/cm$^{-3}$. Once (in the pre-collapse star) the two toroidal structures were formed, then the Torus charge may have survived after the collapse. Thus, currents with unbelievable high values and zero resistivity would produce the Magnetar's fast dynamo action; the outcome may be even a value of $10^{11}$ T. As the time proceeds, the "Torus charge" probably decreases and the circular (azimuthal) current declines, too. In the situation of a magnetar the "Torus structure" with zero resistivity might be the whole spheroid-like layer characterized by $u_e = c'$. This layer may be characterized by a tiny thickness, for instance of $\sim 30$ m. In the Sun's case, the magnetic field close to the tachocline is much stronger than that outside the photosphere; in a magnetar we directly observe the strong magnetic fields since there is no any Convection Zone.

## *4.9. About the Geodynamo*

The weak (and global scale) polar magnetic field of the Sun is mainly produced by the azimuthal currents flowing over the RZ-surface, whereas the strong magnetic fields (directly connected to the surface active regions) are ultimately related to the intense azimuthal currents built up within the two Torus-like structures. In the Geodynamo, where obviously the Torus entity would not be formed, the produced field will probably be the weak polar field related to the outer core of the Earth. That is, strong magnetic fields due to the "fast dynamo" action are not developed in the terrestrial environment, and ***only the slow dynamo action is at work***. A major implication from the latter is probably the aperiodic character of the Geodynamo and its long-lasting processes. The hypothesis that the Earth's magnetic field may reverse when an imbalance of currents occurs in the meridional circuit seems to be a realistic one. For instance, the electrons moving radially at low latitudes are much more than those flowing radially at the very high latitudes. The latter certainly suggests that the solar processes involved in **Fig. 9b and e** would be more or less adopted in the Earth's situation. In this perspective, the solar RZ corresponds to the Earth's core and the Convection Zone to the Earth's mantle. Obviously, the Earth's mantle is not liquid; however, our model is not established on a convection velocity field. Moreover, it is well-known that there are numerous and extreme differences between a planetary and a stellar dynamo; for instance, the mantle conductivity values are much lower than those in the Convection Zone. However, the Earth's meridional current system may be similar to that implemented in the meridional sub-circuit of the "3D solar circuit" (subsection 4.1).

## *4.10. North-south asymmetry in the polarity reversal*

The Sun's polar magnetic fields are much weaker than active region fields; nevertheless, they have far-reaching importance because of their unipolarity over large spatial scales and because of their role in the solar activity cycle [29]. Sporadically a north-south asymmetry in the polarity reversal is observed; that is, the one polar reversal preceded the other for more than a year (e.g., [30, 31]).

In the context of this work, the collapse time of tori close to the equatorial plane defines the onset time for a new activity cycle. In addition, the reverse time of the magnetic field in each hemisphere is determined by the total magnetic field resulting from the Torus current intensity plus the RZ azimuthal current. Therefore, the different toroidal current intensities produce the detected asymmetry of the "polarity reversal time". And the equatorward drift velocity, for each Torus, will be slightly different as depended on the Torus current (look at the subsection 2.14).

## *4.11. The photospheric magnetic field*

The Sun's northward (+) poloidal magnetic component $P_{field}$, as measured on photospheric magnetograms, flips polarity near sunspot cycle maximum, which (presumably) corresponds to the epoch of ***peak internal toroidal currents $T_{current}$***. And the poloidal component peaks at time of sunspot minimum. The toroidal currents flow westward (+) and eastward (-). The cyclic regeneration of the Sun's full large-scale magnetic field and current can thus be thought of as a temporal sequence of the form

$$P_{field}(+) \rightarrow T_{current}(+) \rightarrow P_{field}(-) \rightarrow T_{current}(-) \rightarrow P_{field}(+) \rightarrow \ldots.$$

Toroidal magnetic fields are clearly formed, but the toroidal currents play the chief role.

Obviously our point of view does not comply with the statement that "the dynamo problem can be broken into two sub-problems: generating a toroidal field from a pre-existing poloidal component, and a poloidal field from a pre-existing toroidal component" (e.g., look at the review of Charbonneau, [12]). The above two-fold statement ***mimics the truth***; in reality is a shadow of truth. And the latter may be tightly associated with the old motto that "in the solar dynamo case, the former problem (i.e., the ω-effect) turns out to be easy, but the latter (i.e., the α-effect) is not". In our view the "α-effect" is systematically realized via the (equatorward drifting) two toroidal currents. That is, it is not the statistical outcome achieved from cyclonically rotating magnetic loops; as the latter is typically supposed.

# 4.12. Superconductivity and the zero resistivity in the Sun's Torus

The charged Torus-core region *remains completely unshielded.* The case seems to be no less than miraculous. The electrons behave like the photons in a coherent laser beam. A rough estimate of the Torus supercurrent gives a value of ∼**250 TA**, if the sunspot field is assumed being 1500 G and $\mu = \mu_o$. In this subsection a hypothesis has been formed, in the context of this work, by speculating over the topic of superconductivity.

Certainly, the Torus "superconductivity" is not based on the concept that the electrons change their behaviour from fermions into bosons. However, could a "Cooper pair" of electrons be created in a superconductive material on the basis of the main idea suggested in this work (for the Torus region of the Sun)? A unified theory will probably be a better model approaching phenomena in Cosmos. *The attractive magnetic force*, resulting from the electron or ion (translational) velocities, *may be the driver leading to superconductivity*. Once "Cooper pairs" are formed, finally a macroscopic coherent and rigid wavefunction might be produced. We shall especially deal with the Yttrium Barium Copper Oxide (YBCO) superconductors included in the family of "superstripes". Information relative to this subsection there is in every relative textbook (e.g., look at the books edited by Khare, [32]; Bennemann and Ketterson, [33]; Luiz, [34]).

Example of high-$T_c$ cuprate superconductors includes YBCO, which is famous as the first material to achieve superconductivity above the boiling point of liquid nitrogen. $YBa_2Cu_3O_7$ is a 90 K, type-II superconductor, whereas $YBa_2Cu_3O_6$ displays semiconducting behaviour. The unit cell of $YBa_2Cu_3O_7$ consists of three pseudocubic elementary perovskite unit cells. Each perovskite unit cell contains a Y or Ba atom at the center: Ba in the bottom unit cell, Y in the middle one, and Ba in the top unit cell. Thus, Y and Ba are stacked in the sequence Ba–Y–Ba along the c-axis. All corner sites of the unit cell are occupied by Cu. The structure has a stacking of different layers: $CuO-BaO-CuO_2-Y-CuO_2-BaO-CuO$. One of the key features of the unit cell of $YBa_2Cu_3O_7$ is the presence of two layers of $CuO_2$ with superconductivity taking place between these layers.

The role of the Y layer is to serve as a spacer between the two $CuO_2$ planes and is referred to as "*the charge reservoir*". The Y plane appears to be largely responsible for providing charge carriers to the $CuO_2$ planes. It also determines the degree of anisotropy in the individual high-$T_c$ compounds, as c-axis transport is primarily determined by this layer. The electrical conduction is highly anisotropic, with a much higher conductivity parallel to the $CuO_2$ plane than in the perpendicular c-direction. The value of the coherence length is found to be highly anisotropic. The coherence length parallel to the c axis (i.e., perpendicular to the $CuO_2$ plane) is typically 2–5 Å, and in the ab plane, the value is typically 10–30 Å. The atomic diameter of the Yttrium atom is 3.8 Å. In conventional low-$T_c$, type I superconductors, the coherence length is 1000 Å, which is several orders of magnitude larger than that in high-$T_c$ superconductors.

Taking into account all the exhibited information, a preliminary suggestion may be as follows: We assume that the applied external electric field accelerates electrons (over the $CuO_2$ planes), at the flanks of the high conductivity channel, leading finally to paired electrons (or paired holes, if we assume mobile holes moving with opposite spin). That is, we emphasize on the role potentially played by the two elementary homo-parallel electron currents; we focus on the magnetic force effect caused by the two moving electrons. As a matter of fact, the electrical currents in the $YBa_2Cu_3O_7$ superconductor are carried by holes induced in the oxygen sites of the $CuO_2$ sheets.

In **Fig. 17**, two electrons move in parallel, while remaining in adjacent $CuO_2$ planes and hoping from one lattice site (marked with a black solid circle) to another. Each electron is subject to attractive and repulsive forces. First, there is an attractive electrostatic force $\mathbf{F_1}$ from the neighbouring positively charged Yttrium ion; second, a repulsive electrostatic force $\mathbf{F_2}$ from the other (paired) electron, and third, an attractive magnetic force $\mathbf{F_3}=\mathbf{JXB}$ caused from the two elementary current elements. Under these circumstances, the attractive force may be greater than the repulsive one (i.e., $F_1+F_3>F_2$), leading to a "Cooper pair". Consequently, the value of velocity, for an electromagnetic wave perpendicular to the $CuO_2$ plane, is particularly important.

It should be underlined that the current J, flowing over a $CuO_2$ plane, attracts all the charge carriers being located close to the other $CuO_2$ plane. Thus, we practically have two currents flowing in parallel over the $CuO_2$ layers.

Like high-$T_c$ superconductors made of cuprate ceramics, $MgB_2$ is an intermetallic (layered) superconductor (with $T_c$=40 K); while undoped cuprates are insulators at ordinary temperatures, however, $MgB_2$ is always a metal. It consists of hexagonal honey-combed planes of boron atoms separated by planes of magnesium atoms, with the magnesiums centered above and below the boron hexagons. The mechanism at work may be roughly similar to what is previously hypothesized.

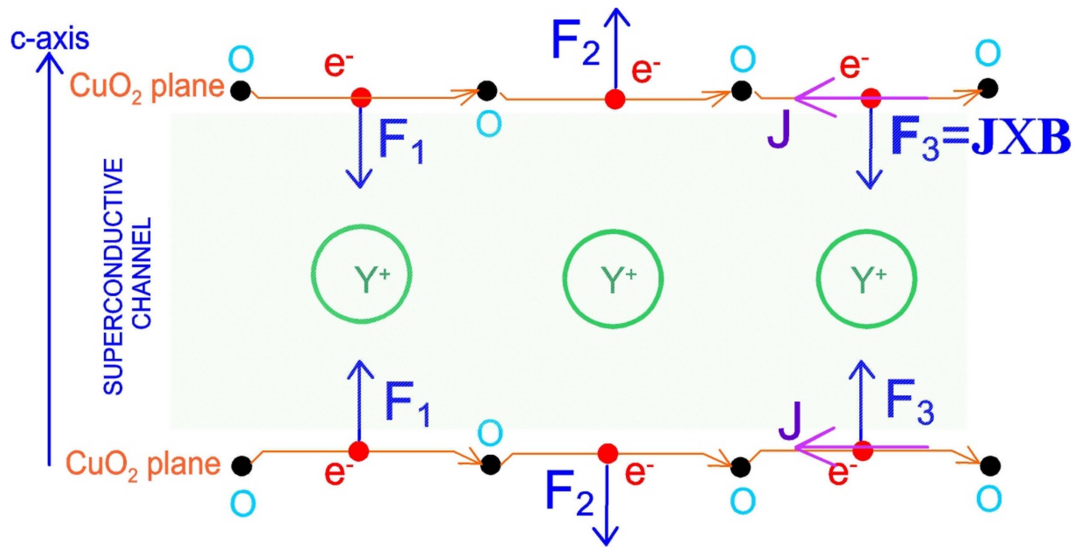

**Fig. 17.** Two electrons forming a "Cooper pair" move in parallel over two adjacent $CuO_2$ planes. Y stands for a (positively charged) Yttrium atom and O for a lattice site. In each electron hop, the following three forces are involved: First, an attractive force, F1, between an electron and an Yttrium atom; second, a repulsive electrostatic force, $F_2$, from the mutually repelled electrons, and third, an attractive magnetic force, $F_3$=JXB, produced from the two elementary current elements. Under these circumstances, the attractive force may be greater than the repulsive one (i.e., $F_1+F_3>F_2$), leading to a "Cooper pair". The layers of $CuO_2$ causes a large anisotropy in normal conducting and superconducting properties, since electrical currents are carried by holes induced in the oxygen sites of the $CuO_2$ sheets.

## 4.13. *Speculation about an "electron degeneracy pressure" in the Torus core region*

One may speculate that an electron-degenerate matter may have developed within the negatively charged Torus. And degenerate matter has a pressure much higher than the partial pressure of ions. Adding electrons into the Torus may force the electrons into higher-energy quantum states. This requires a compression force which is made manifest as a resisting pressure and the degeneracy pressure is depended on the density of electrons. Outside the Torus, the thermal pressure dominates so much that degeneracy pressure can be ignored. In "white dwarf stars" the supporting force comes from the degeneracy pressure of the electron gas in their interior; however, in the Torus situation, the electron degeneracy pressure would not be related to any gravitational collapse. It might be a pressure built up against the pressure from the magnetic force attracting additional electrons into the Torus. The magnetic force increases as the Torus current enhances, and the enhanced current further intensifies the magnetic force.

Finally, a negatively charged Torus, with its exotic character, might be termed as a "negatively overcharged electron trap"; perhaps a particular plasma category in Cosmos. In addition, the Torus material has probably developed "the property of superconductivity" at extremely high temperatures (i.e., ~$2\cdot10^6$ K). However, an electron degeneracy pressure within the Torus is not produced. In the next sunspot cycle the Torus will be positively overcharged. Therefore, the overcharged core Torus region should not be related to any degenerate matter.

# *5 . Conclusion*

We propose a "fast dynamo" action potentially actuated in the Sun's case and serving, as a model, for similarly arranged "stellar dynamos". The key process of the suggested dynamo is not supposed to be any convection "velocity field" in the Convection Zone of the Sun, but a peculiar "superconductive conduction path" developed within two Torus-like structures. Each Torus is characterized by extremely high temperatures (as formed within the tachocline) and its toroidal current flows under zero resistivity conditions. ***This exotic property results from the fact that a net charge is accumulated inside the Torus.*** A charge that alternates from positive to negative, or vice versa, in two successive sunspot cycles. The produced currents ultimately set the 11-year sunspot cycle and reverse the magnetic field polarity within the whole 22-year solar cycle. In the case of a negatively charged Torus, the local speed of an electromagnetic wave, $c'$, is comparable with the velocity of electrons, $u_e$, being the local rotational tangential speed of the Sun, $u_\phi$. Therefore, under this condition, ***two electrons do not always repel each other*** (which happens when $u_e < c'$); moreover, they can remain unaffected ($u_e = c'$) or they are even mutually attracted ($u_e > c'$). That is, the plasma is "Debye shielded", "Debye anti-shielded" or the particle distribution remains completely unaffected. The zero resistivity accomplishes the "fast dynamo" action. There are numerous and serious consequences of the proposed model. Within the context of this work, we extensively discuss issues covering topics as the mechanism via which magnetic loops may be emerging out of the photosphere, the solar flares, the CMEs and the associated issue of energetic electron acceleration.

Finally, we would like to stress that if the suggested processes (i.e., of "charge separation and net charge confinement" within the two tori) were really implemented in nature, then any further attempt to resolve the Sun's dynamo problem based on the MHD approach, and specifically through the "induction equation" along with a prescribed plasma convection velocity field, would inevitably fail. In contrast, if the presently suggested model were valid offering a realistic perspective, and the main exhibited aspects were acceptable, then many longstanding unresolved questions concerning the powerful CMEs, the flares, the helicity generation and the stellar dynamo could be readily addressed. This author thinks that the major outcome of this work is that the magnetic fields alone, without taking into account the charge distributions and currents, could not cause a forward movement helping us to resolve the enigmatic behaviour of cosmic plasmas. Certainly, any proposed model can only describe the cosmic phenomena in an approximate manner, since the universe is fluently changing based on various unexpected constellations and unpredictable parameters. Eventually, cosmos may always be based on a novel and unanticipated logical structure.